\documentclass[epj]{svjour}
\usepackage{graphics,color}
\usepackage[spaces,hyphens]{url}
\usepackage{hyperref,breakurl}
\newcommand{\dif}{\mathrm{d}}
\newcommand{\snn}{\sqrt{s_{\mathrm{NN}}}}
\newcommand{\ie}{\emph{i.e. }}
\newcommand{\eg}{\emph{e.g. }}

\begin{document}
\title{Particlization in hybrid models}

\author{Pasi Huovinen\inst{1} \and Hannah Petersen\inst{2}
}                     
\institute{Institut f\"ur Theoretische Physik, 
     Johann Wolfgang Goethe-Universit\"at,
     D-60438 Frankfurt am Main, Germany
           \and 
    Department of Physics, Duke University, Durham, NC 27708-0305, USA }
%
\date{}
%

\abstract{In hybrid models, which combine hydrodynamical and transport
  approaches to describe different stages of heavy-ion collisions,
  conversion of fluid to individual particles, particlization, is a
  non-trivial technical problem. We describe in detail how to find the
  particlization hypersurface in a 3+1 dimensional model, and how to
  sample the particle distributions evaluated using the Cooper-Frye
  procedure to create an ensemble of particles as an initial state for
  the transport stage. We also discuss the role and magnitude of the
  negative contributions in the Cooper-Frye procedure.
} 
\maketitle
\section{Introduction}
\label{intro}

In recent years the so-called hybrid
models~\cite{Bass:2000ib,Teaney:2001av,Hirano:2005xf,Nonaka:2006yn,Petersen:2008dd,Werner:2010aa,Song:2010mg,Karpenko:2012yf,Hirano:2012kj}
have become very popular in describing the expansion and evolution of
the hot dense matter created in ultrarelativistic heavy-ion
collisions. In these models the different stages of the expansion are
described using different models: The early stages of the expansion,
when the matter is presumably partonic, are described using ideal or
dissipative fluid dynamics, whereas the late dilute stage after
hadronization is described using a hadronic transport model like
RQMD~\cite{Sorge:1995dp}, UrQMD~\cite{Bass:1998ca} or
JAM~\cite{Nara:1999dz}.

Hybrid models are conceptually attractive since they attempt to
combine the best features of fluid dynamics and transport: The
complicated microscopic dynamics of hadronization can be sidestepped
by assuming it to happen adiabatically, in which case it can be
described using fluid dynamics. On the other hand, the dilute hadronic
matter is assumed to be highly dissipative and deviate gradually from
local equilibrium. Description of such a matter using fluid dynamics
is demanding, but is trivial using microscopic transport, which by
construction can handle matter arbitrarily far from equilibrium. Also,
the transport models describe the last rescattering of particles,
so-called freeze-out, based on individual scattering cross
sections. Thus freeze-out is not controlled by a parameter which value
is known only after comparison with data.

Furthermore, it has been realised that a consistent comparison with
the data is easier if one generates an ensemble of particles
event-by-event, and analyses these ensembles in a similar way than the
actual data has been
analysed~\cite{Hirano:2012kj,Holopainen:2010gz,Petersen:2010cw}. In
hybrid models this requires no further effort since the transport
models are based on propagation of individual particles along their
semiclassical trajectories, and thus require the generation of such
ensembles to begin with.

However, connecting two approaches with very different degrees of
freedom --- densities, velocity, and possible dissipative currents in
the fluid, and individual particles in the cascade --- is highly
nontrivial. Instead of running hydro and cascade side by side and
using the cascade calculation to provide the boundary conditions for
fluid dynamics, all the present hybrid models solve fluid dynamics
independently of the cascade taking the boundary condition as vacuum
at infinity (see discussion in Ref.~\cite{Rischke:1998fq}). The
boundary where one switches from fluid dynamical to transport
description is then determined a posteriori, once the evolution of the
fluid is known. This boundary, or switching surface, is usually chosen
to be a surface of constant temperature, energy density, or time. The
particle distributions on this surface are evaluated using the
Cooper-Frye procedure~\cite{Cooper:1974mv}. These distributions are
sampled to generate an ensemble of particles with well defined
positions and momenta, and these ensembles are used as an initial
state for the transport stage. Note that even if the transport stage
describes the final freeze-out of the particles without additional
parameters, the choice of the criterion where to switch from fluid to
cascade is an equally free parameter than the freeze-out temperature
in a conventional hydrodynamical model. Unfortunately the final
results are also somewhat sensitive to this switching criterion as
discussed in Ref.~\cite{Hirano:2012kj}, and as we will discuss later.

We want to emphasise that this switching from fluid dynamics to
transport is not freeze-out. By definition, there are no
rescatterings, only resonance decays after freeze-out. Thus, if one
switches from fluid to transport at freeze-out, the transport stage is
unnecessary. It does nothing else but lets the resonances decay. On
the other hand, the switching is not necessarily hadronization
either. Since the deconfinement transition is a smooth crossover,
there is no clear point where hadronization should take place, and it
is easier to choose to switch at a stage where the system has already
hadronized. In the following we assume that there is no change in the
physics nor in the properties of the system on the switching
surface. It is only a change in the description of the system. Thus we
call the process of changing from fluid dynamics to transport
particlization, conversion of fluid to particles.

In this paper we describe some technical aspects of particlization in
hybrid models, and how different choices of the switching criterion
and constraints imposed to the sampling of the distributions affect
the final particle distributions in $E_{\rm lab}=160A$ GeV Pb+Pb (SPS)
and $\snn=200$ GeV Au+Au (RHIC) collisions. In particular we discuss
finding the particlization surface and its properties in
section~\ref{sec:surface}, the negative contributions of Cooper-Frye
procedure and how to take them into account when sampling the
distributions in section~\ref{sec:negative}, the actual sampling
procedure in section~\ref{sec:sampling}, and calculations within one
specific hybrid approach in section~\ref{sec:results}.

\section{Surface finding}
  \label{sec:surface}

The Cooper-Frye procedure~\cite{Cooper:1974mv} for calculating
particle distributions on a surface is based on evaluating a particle
four-current through a surface, and a kinetic theory decomposition of
a four-current $j^\mu$ in terms of particle distribution $f(p,x)$:
\begin{eqnarray}
 \label{eq:CF}
  N & = & \int_\sigma \dif\sigma_\mu j^\mu(x)           \nonumber
        = \int_\sigma \dif\sigma_\mu\!\int \frac{\dif^3 p}{E}\, p^\mu f(x,p) \\
 \Longrightarrow
 E\frac{\dif N}{\dif p^3} & = & \int_\sigma \dif\sigma_\mu p^\mu f(x,p)
                         \approx \sum_\sigma \Delta\sigma_\mu p^\mu f(x,p).
\end{eqnarray}

Thus one needs to find not only the location of the surface $\sigma$
where one applies the Cooper-Frye formula, but also its normal. If we
knew the analytic expression for the surface, its normal would be
simply given by~\cite{Hung:1997du,Misner}
\begin{equation}
 \dif\sigma_\mu = \varepsilon_{\mu\alpha\beta\gamma}
                  \frac{\partial x^\alpha}{\partial a}
                  \frac{\partial x^\beta}{\partial b}
                  \frac{\partial x^\gamma}{\partial c}
                  \dif a\, \dif b\, \dif c\, ,
\end{equation}
where $a,b,c$ are the coordinates on the surface and
$\varepsilon_{\mu\alpha\beta\gamma}$ is the totally antisymmetric
Levi-Civit\`a tensor. However, we do not have an analytic expression
since we solve the evolution of the system numerically. Finding the
location of an isosurface on a discrete grid is easy, but evaluating
the size\footnote{The size enters as the length of the discrete normal
  vector $\Delta\sigma_\mu$.} and normal of each discrete surface
element is not, since one has to make sure that the surface elements
cover the entire surface without leaving holes or double counting any
part of the surface.

In computer graphics and image processing the problem of finding an
isosurface of a discrete scalar field is well known, and many
algorithms have been proposed for the task, see, \eg
Ref.~\cite{Roberts}, and references therein. One of the best known of
these algorithms is so called Marching Cubes
algorithm~\cite{Lorensen}, a simplified version of which was
implemented in a hydrodynamical model by Kataja and
Venugopalan~\cite{Venugopalan:1994ux}, and subsequently used in
AZHYDRO~\cite{Kolb:2000sd}. The original Marching Cubes algorithm was
extended for hypersurfaces in four dimensional space in
Ref.~\cite{Roberts}, and a simplified version implemented in a 3+1
dimensional hydrodynamical model by Schenke~\cite{Schenke:2010nt}.
However, it is known that the original Marching Cubes algorithm cannot
resolve all possible surface configurations, and may leave holes in
the final surface~\cite{Duurst,Chernyaev}. This problem is not
serious: As quoted in Ref.~\cite{Schenke:2010nt}, only 1\% of the
surface elements in a typical heavy-ion collision calculation are not
fully resolved. But, if one is doing event-by-event hydrodynamical
calculations with irregular initial conditions, one may expect much
more complicated structures to appear, and the Marching Cubes
algorithm may leave more holes in the surface. To avoid this problem
altogether, we have slightly modified the algorithm proposed in
Ref.~\cite{Wyvill}, and generalised it for finding a three dimensional
hypersurface in four dimensional space\footnote{Fortran and C++
  subroutines, cornelius, implementing this algorithm in 3D and 4D,
  are available at
  \url{https://karman.physics.purdue.edu/OSCAR/}}. For the lack of a
better name, we call it here ``disordered lines'' algorithm. This
algorithm may also have been implemented in Ref.~\cite{Hung:1997du},
but the description there is too vague to tell.

In algorithms solving the equations of motion of fluid dynamics like
SHASTA~\cite{Boris73}, a grid point is considered to be in the middle
of a corresponding volume element. For purposes of surface finding it
is useful to consider the dual of the grid, where the grid points are
thought to be at the corners of a volume element, and the values
within a volume element can be obtained by interpolation. To find the
location of the surface it is practical to ``march'' through the
entire grid and check every volume element whether the values at the
corners are all above or all below the isovalue. If not, \ie if some
of the corners are above, and some below the isovalue, the surface
passes through this volume element.

An alternative to this exhaustive search method is the continuation
method~\cite{Bloomenthal}, which after finding a surface element,
checks only the neighbouring volume elements which one of them
contains the surface, moves to that element, and continues until it
either finds the edge of the grid or returns to the volume element it
found first. The continuation method is faster than the exhaustive
search, but if the surface is disjoint, it will not necessarily find
all parts of the surface, whereas the exhaustive search method by
construction finds all parts of the surface.

\begin{figure}
\resizebox{0.5\textwidth}{!}{
  \includegraphics{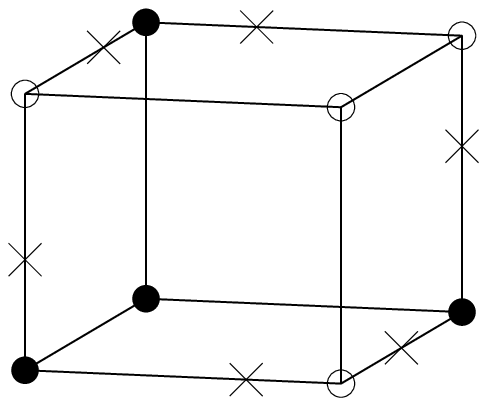}
  \includegraphics{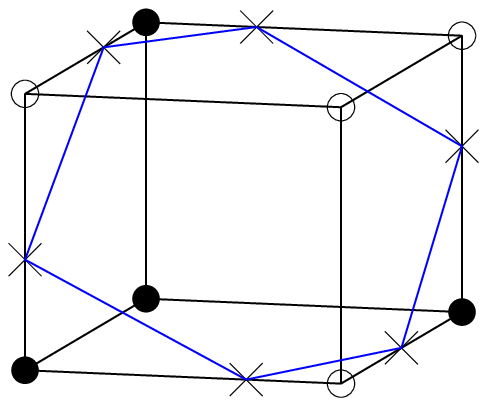}
}
\caption{A 3D volume element, a cube, with four corners above (solid
  dots) and four below (open dots) the isovalue. In the left panel the
  points where the interpolated value is the isovalue are marked with
  crosses. In the right panel these intersection points have been
  connected to form a polygon. This polygon is a surface element of
  the isosurface.}
\label{fig:cube}
\end{figure}

In the following we call volume elements (hyper)cubes, and explain the
algorithm in a 3D case first. Once a cube containing the isosurface is
found, the positions of the isosurface on the edges of the cube can be
found by linear interpolation, see illustration in Fig.~\ref{fig:cube}. 
How to sort these intersection points to form a polygon (or a
polyhedron in 4D) which does not cross itself is the crucial part of
any algorithm. It can be done in three different ways~\cite{Schlei}:
By using protomeshes~\cite{Schlei:2010yn}, by using a lookup table
like in the original Marching Cubes
algorithm~\cite{Roberts,Lorensen,Chernyaev}, or
algorithmically~\cite{Bloomenthal,Thirion,Fidrich,Pang:2012he} like we
do here.

\begin{figure}
 \begin{center}
  \resizebox{0.35\textwidth}{!}{\includegraphics{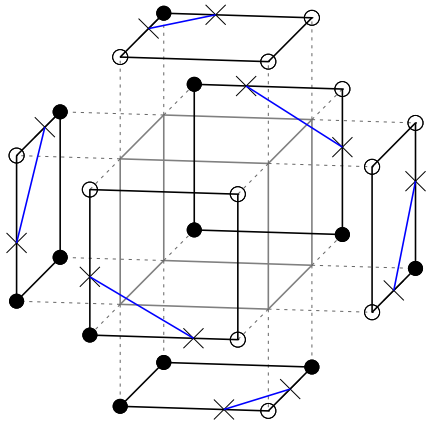}}
 \end{center}
\caption{Reduction of a three dimensional
  problem into a series of two dimensional problems: We find the
  surface element by looking for its edges on the faces of the
  cube.}
\label{fig:facebyface}
\end{figure}

\begin{figure}
\resizebox{0.5\textwidth}{!}{
  \includegraphics{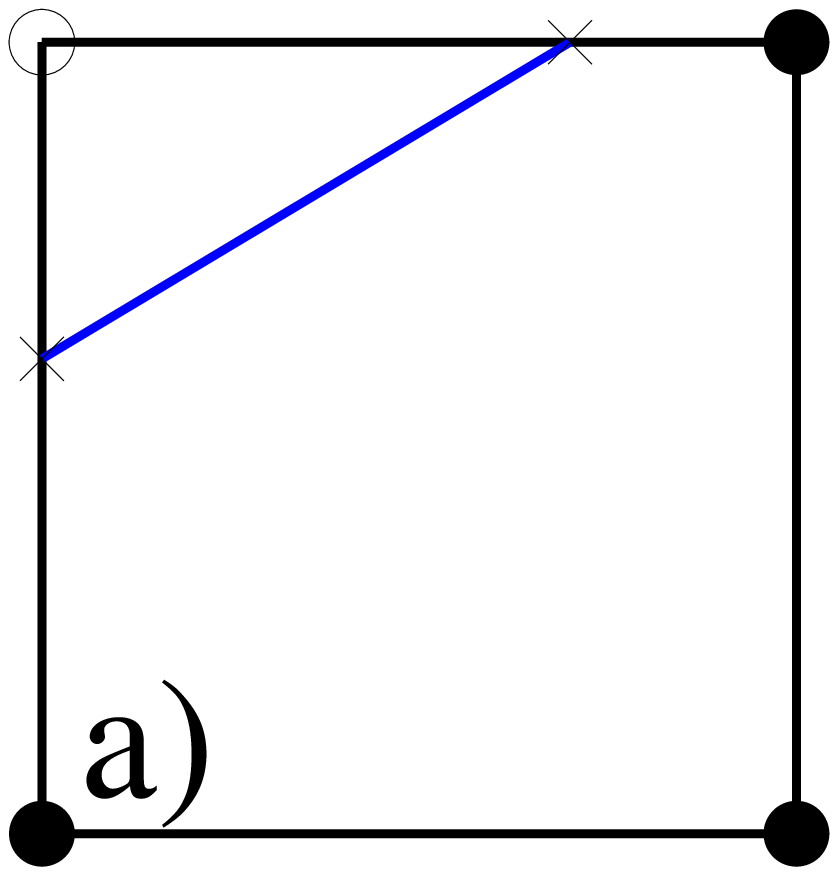}
  \includegraphics{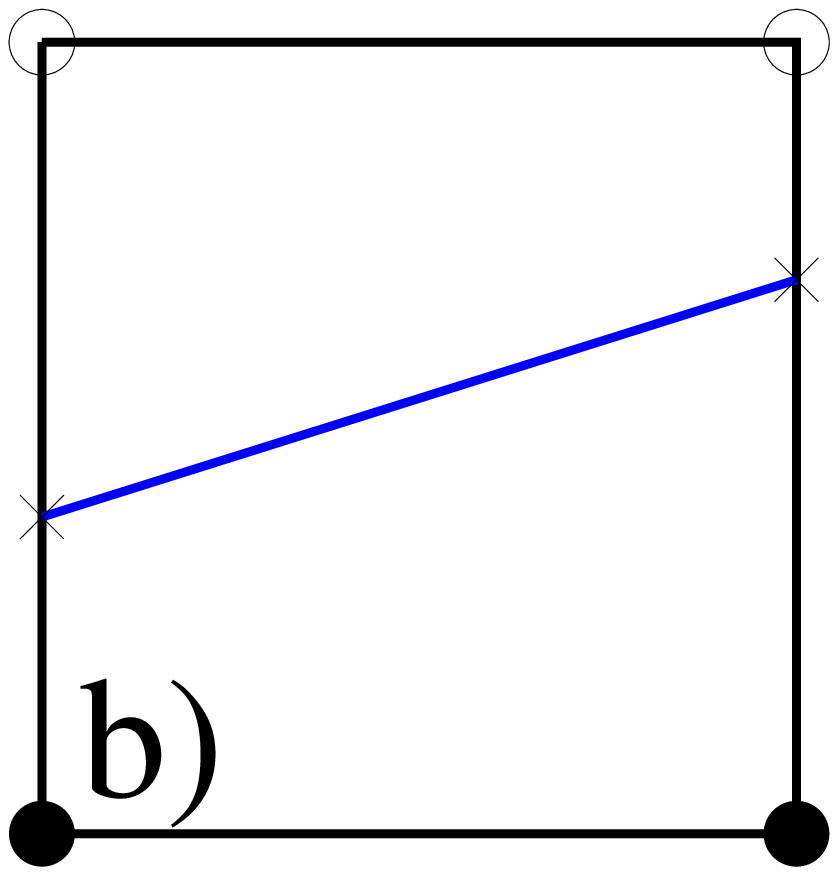}
  \includegraphics{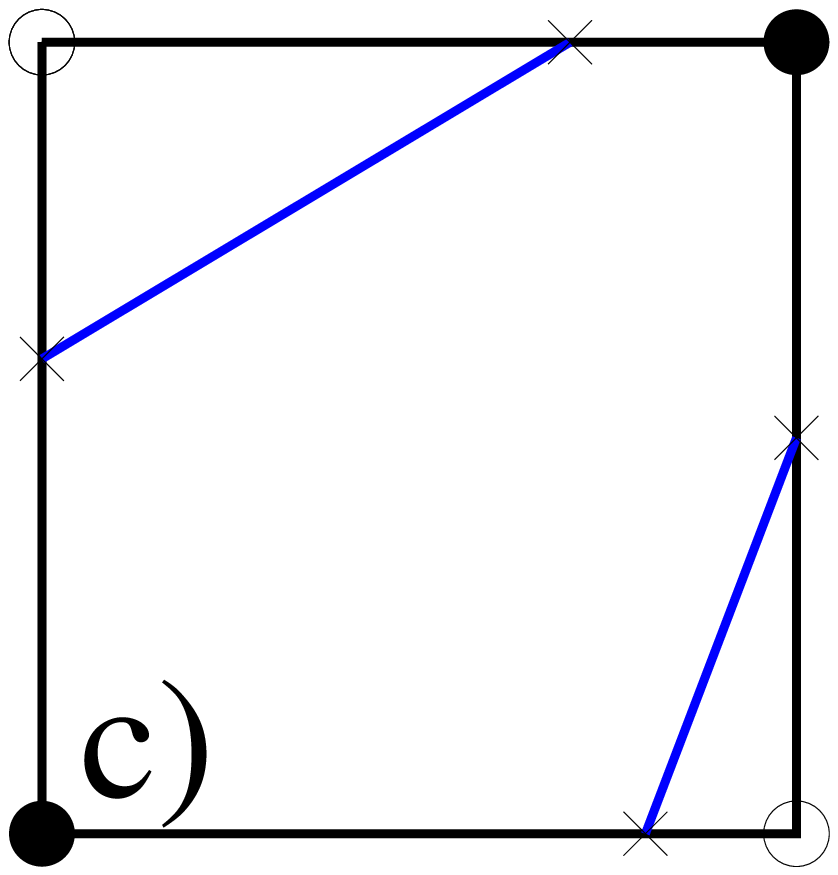}
  \includegraphics{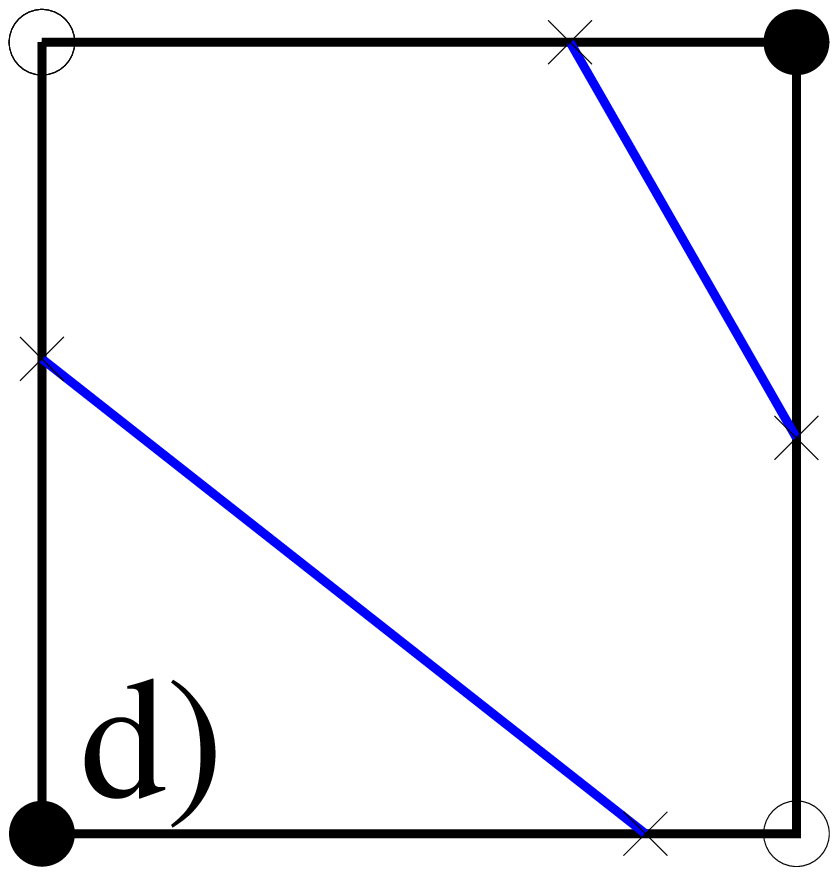}
}
\caption{The nontrivial cases of connecting intersection points at the
  edges of the face.}
\label{fig:squares}
\end{figure}

In three or more dimensions the ordering of the intersection points is
difficult because of the many possibilities involved, but in two
dimensions it is almost trivial. Therefore we reduce the problem of
finding the surface into a series of two dimensional problems of
finding a line within a square, a line which is an edge of the polygon
we want to find, see Figs.~\ref{fig:cube} and~\ref{fig:facebyface}. If
all of the corners of a square are above or below the isovalue, there
is of course no edge of the polygon there. If only one of the corners
is above or below, we have the configuration shown in
Fig.~\ref{fig:squares} a), and if two neighbouring corners are above
the isovalue, the configuration is shown in Fig.~\ref{fig:squares} b).
The case where two corners above the isovalue are located diagonally
is ambiguous. How to choose between configurations shown in
Figs.~\ref{fig:squares} c) and d)?  We follow the original idea of
Ref.~\cite{Wyvill}: If the value interpolated at the center of the
square is above or equal to the isovalue, the center must be inside of
the surface, and if it is below, the center must be outside of the
surface. A more sophisticated rule has been suggested in
Refs.~\cite{Chernyaev,Thirion}: The values on the face are
interpolated bilinearly. The isocontours in a bilinear interpolation
are hyperbolas, and thus one should interpolate the value at the
center of this hyperbola to decide whether the corners above or below
the isovalue are linked.  A special case occurs when the value in any
of the cube corners is the isovalue. In that case we consider such a
corner to be within the surface, and place intersection points
$10^{-9}$ times the edgelength away from the corner, on those edges of
the cube where the neighbouring corner has a value smaller than the
isovalue.

\begin{figure}
 \begin{center}
  \resizebox{0.4\textwidth}{!}{\includegraphics{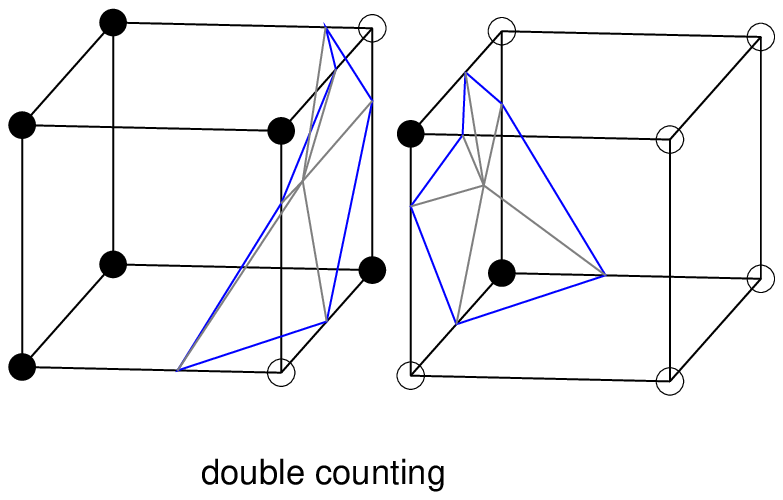}}
  \resizebox{0.4\textwidth}{!}{\includegraphics{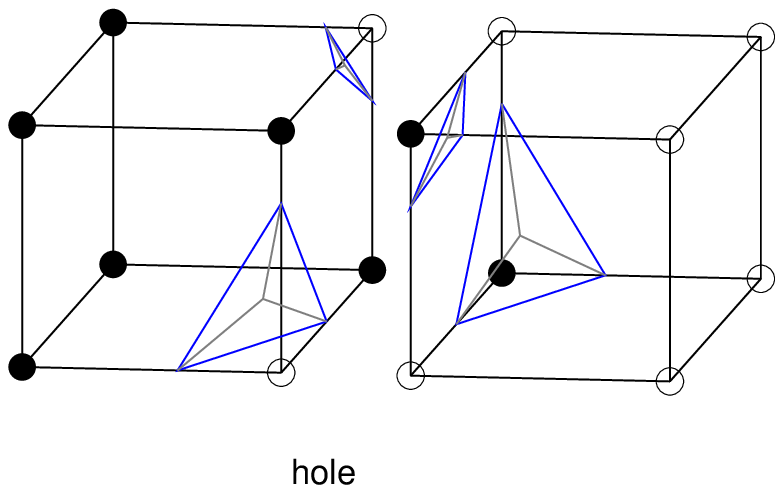}}
  \resizebox{0.4\textwidth}{!}{\includegraphics{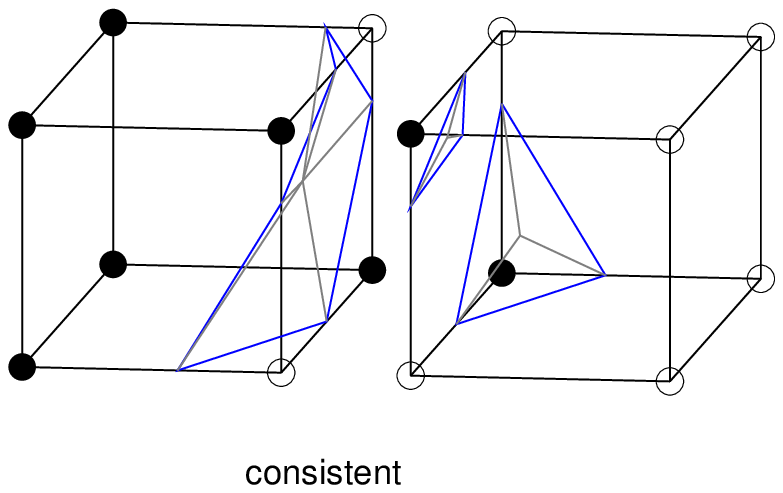}}
 \end{center}
 \caption{An example of a configuration where badly resolved ambiguity
   may lead to double counting or a hole on the surface. The cubes
   left and right represent neighbouring cubes separated for the sake
   of clarity.}
\label{fig:hole}
\end{figure}

After the intersection points at each face have been found, each edge
of the polygon is characterised by two points: its ends. An important
difference between our algorithm and that of
Refs.~\cite{Wyvill,Thirion} is that in most common cases we do not
need to order the edges of the polygon so that the next edge in the
list begins where the previous one ends. We may check the faces of the
cube in arbitrary order, and keep the edges of the polygon stored in
the same arbitrary order. Important exceptions to this rule are the
cases where the face by face search returns six or more edges: The
surface may consists of two or more disconnected parts, see
Fig.~\ref{fig:hole}, and we have to sort the edges in sequence to find
out whether the surface is disjoint, and which edges belong to which
polygon. We group the edges accordingly, and treat the separate
polygons independently of each other.

Note that allowing the surface to consist of several parts is
necessary for the consistency of the surface. Fig.~\ref{fig:hole}
depicts a case when a badly resolved ambiguity on a face may lead to a
hole on the surface, or counting parts of the surface twice. Since we
solve the ambiguities by using only the information available at the
face of a cube, the face is always resolved in the same way, no matter
in which cube it is taken to belong to. Thus the surface elements form
a consistent surface without holes nor double counted elements.
Furthermore, the rules we described are sufficient to resolve any
configuration of values at the cube corners.

\begin{figure}
 \begin{center}
  \resizebox{0.5\textwidth}{!}{\includegraphics{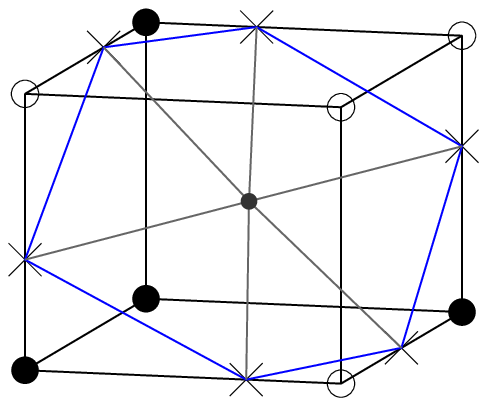}
                               \includegraphics{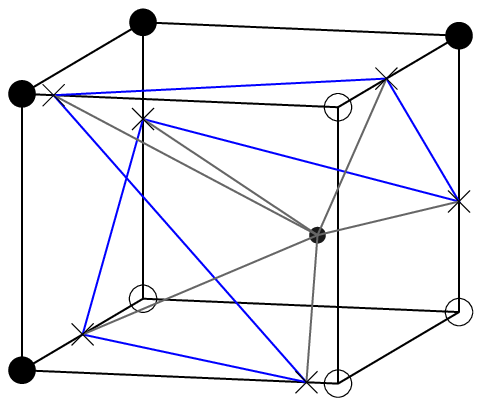}}
 \end{center}
\caption{Examples of triangularisation of the polygon in a simple and
  a complicated case.}
\label{fig:triangles}
\end{figure}

After the possible ordering of the edges and dividing them in separate
groups if they form two or more disconnected surface elements, we
evaluate the area and normal of the polygon(s). The use of Cooper-Frye
procedure requires not only the surface area and normal, but also the
values of the hydrodynamical fields on the surface. They are best
evaluated as interpolated values at the centroid of the surface. Since
we have to evaluate the centroid for this purpose anyway, we
triangularise the polygon using the centroid: By connecting the ends
of each edge of the polygon to the centroid, we obtain a set of
triangles which cover the entire polygon. As depicted in
Fig.~\ref{fig:triangles}, the polygon is not necessarily planar. A sum
of the areas and normal vectors of the triangles approximates the area
and normal of the polygon as
\begin{equation}
  \Delta\vec{\sigma} = \sum_i \frac{1}{2}f_i\mathbf{a}_i\times\mathbf{b}_i,
\end{equation}
where $\mathbf{a}_i$ and $\mathbf{b}_i$ are vectors from the centroid
to the ends of the $i$th edge of the polygon, and $f_i = \pm 1$ is
chosen so that each of the normal vectors
$\frac{1}{2}f_i\mathbf{a}_i\times\mathbf{b}_i$ is directed towards
lower values, \ie outside.

\begin{figure}
 \begin{center}
  \resizebox{0.3\textwidth}{!}{\includegraphics{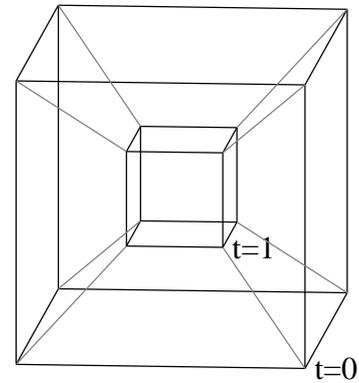}}
 \end{center}
\caption{A 2D projection of a 4D hypercube. In the figure, the cube in
  the middle is the hyperface of the hypercube at coordinate $t=1$,
  the large cube around it is the hyperface at $t=0$, and the grey
  lines connecting them are edges with constant values of coordinates
 $x,y,z$.}
\label{fig:4D}
\end{figure}

\begin{figure}
 \begin{center}
  \resizebox{0.4\textwidth}{!}{\includegraphics{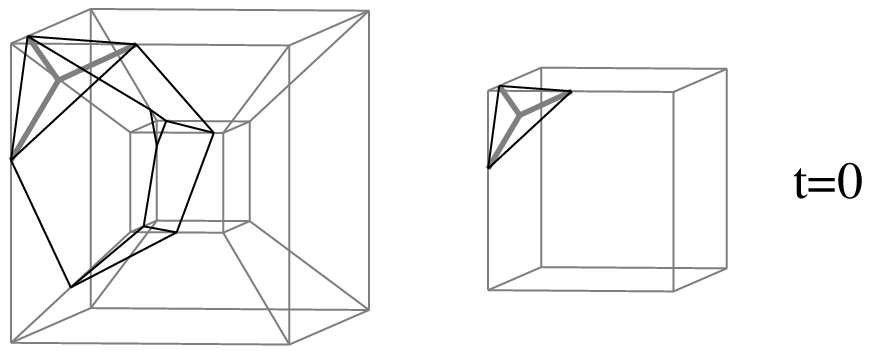}}
  \resizebox{0.4\textwidth}{!}{\includegraphics{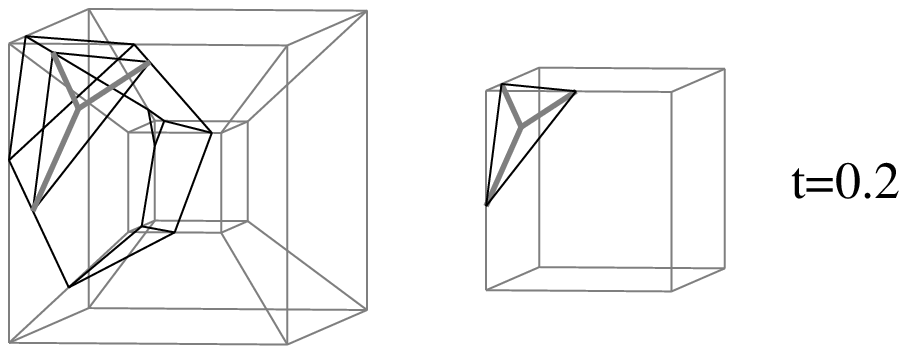}}
  \resizebox{0.4\textwidth}{!}{\includegraphics{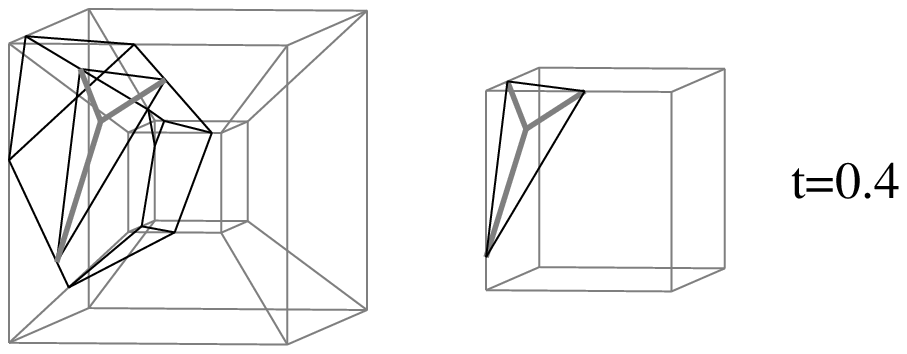}}
  \resizebox{0.4\textwidth}{!}{\includegraphics{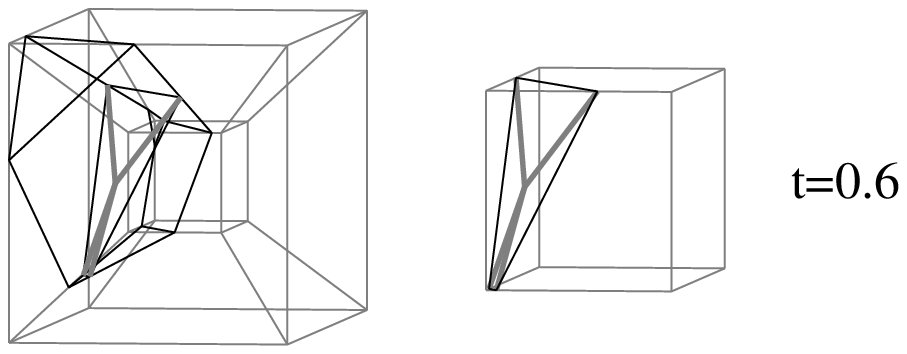}}
  \resizebox{0.4\textwidth}{!}{\includegraphics{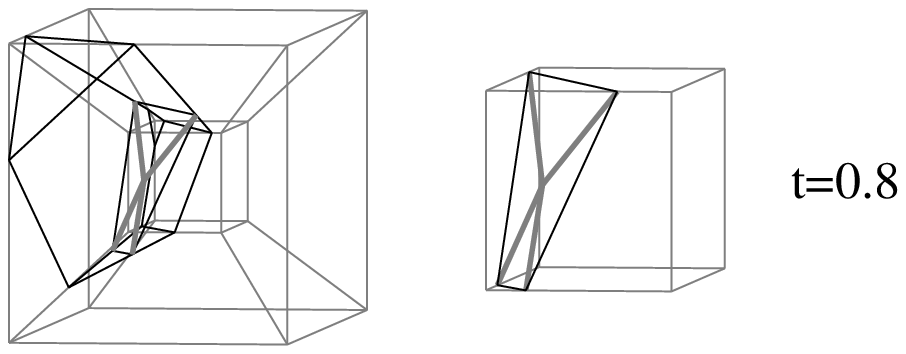}}
  \resizebox{0.4\textwidth}{!}{\includegraphics{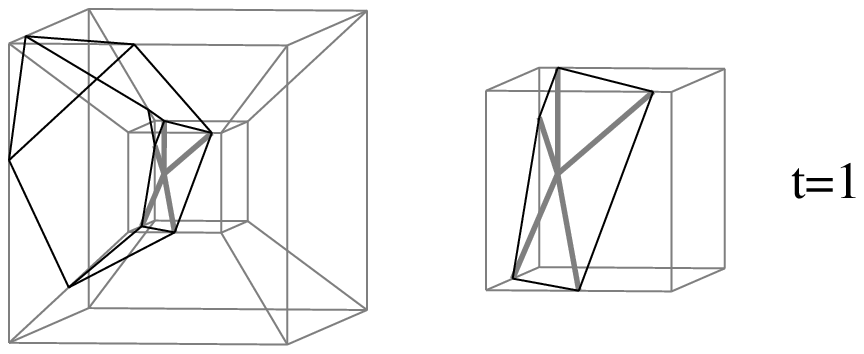}}
 \end{center}
 \caption{Time evolution of a surface element in four (left) and three
   dimensional (right) representation of a volume element. A 2D
   polygon moving in time spans a 3D polyhedron in a hypercube.}
\label{fig:evo}
\end{figure}

\begin{figure}
 \begin{center}
  \resizebox{0.35\textwidth}{!}{\includegraphics{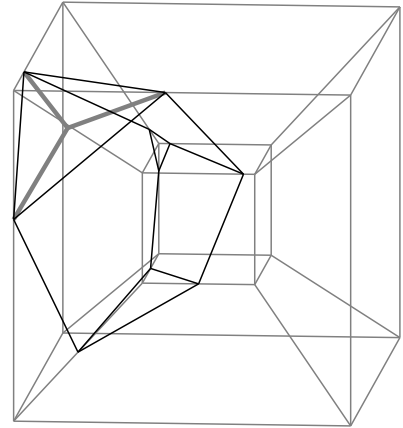}}
 \end{center}
 \caption{A hypersurface element, a polyhedron, within a hypercube. A
   triangularisation of the polyhedron's face with $t=0$ is shown.  We
   do not show the triangularisation of the other faces for the sake
   of clarity.}
\label{fig:polyhedron}
\end{figure}

To make the illustrations of the hypersurfaces in a hypercube more
understandable, we first show a simple 2D projection of a 4D hypercube
in Fig.~\ref{fig:4D}, and how an evolving 2D surface in 3D space spans
a 3D hypersurface in 4D spacetime in Fig.~\ref{fig:evo}. This
hypersurface element forms a polyhedron within a hypercube
(Fig.~\ref{fig:polyhedron}), and like a polygon can be divided into a
group of triangles, a polyhedron can be divided into tetrahedra.

\begin{figure}
 \begin{center}
  \resizebox{0.5\textwidth}{!}{\includegraphics{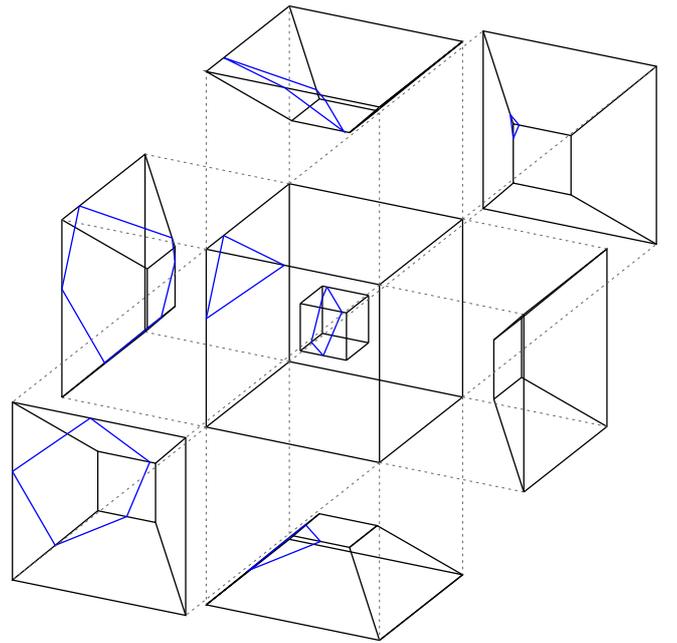}}
 \end{center}
\caption{Reduction of a four dimensional problem into a series of
  three dimensional problems.}
\label{fig:4Dsplit}
\end{figure}

We proceed analogously to the 3D case. We first divide the problem
into eight three dimensional problems: As a cube consists of six
squares, a hypercube consists of eight cubes, see
Fig~\ref{fig:4Dsplit}.  The surface in each of these is found in the
same way than described above. The centroid of these 2D surfaces,
which form the faces of the tetrahedra, is calculated, and the corners
of the triangles forming these faces are recorded. This is illustrated
in Fig.~\ref{fig:polyhedron}, where the triangularisation of the face
within the $t=0$ hyperface is shown.

As in 3D the ordering of the edges is not important unless the surface
consists of several disconnected parts. If the number of edges and the
number of hypercube corners above and below isovalue indicate that
this is possible, we order the edges to group them according to the
polyhedron they belong to, and treat the groups as separate surfaces.

\begin{figure}
 \begin{center}
  \resizebox{0.35\textwidth}{!}{\includegraphics{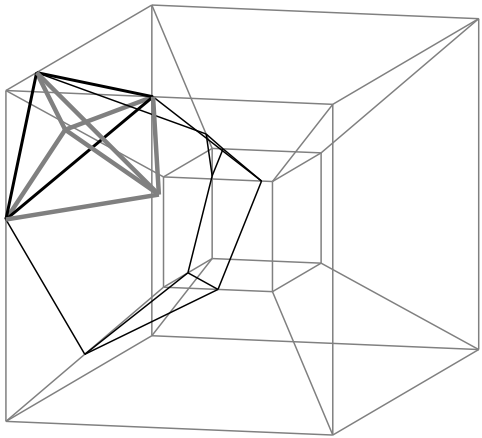}}
 \end{center}
\caption{A part of the tetrahedronisation of the polyhedron based on
  triangularisation of the faces of the polyhedron. Full
  tetrahedronisation is not shown for the sake of clarity.}
\label{fig:tetra}
\end{figure}

We evaluate an approximative centroid for the polyhedron. Analogously
to the 2D surface in a 3D space, connecting the centroid to the
corners of the triangles forming the faces of the polyhedron, creates
a set of tetrahedra which fills the volume of the polyhedron, see
Fig.~\ref{fig:tetra}. The hyperarea, \ie the volume, of the polyhedron
and its normal can be calculated as a sum of the volumes and normals
of its constituent tetrahedra~\cite{Schenke:2010nt}:
\begin{equation}
  \Delta\sigma_\mu = \sum_i \varepsilon_{\mu\alpha\beta\gamma}
                           \frac{1}{6}f_ia_i^\alpha b_i^\beta c_i^\gamma,
\end{equation}
where $a_i$, $b_i$ and $c_i$ are vectors from the centroid to the
corners of the $i$th triangle, and $f_i = \pm 1$ chosen so that each
of the normal vectors of the tetrahedra is directed towards lower
values, \ie outside.

We want to emphasise that this algorithm is by no means the only
approach for creating consistent surfaces with no holes nor double
counting~\cite{Chernyaev,Thirion,Fidrich}. Especially Bernd Schlei's
VESTA and STEVE algorithms~\cite{Schlei:2010yn}, used in
Ref.~\cite{Cheng:2010mm}, are faster than ours and produce even higher
quality surfaces.

In the recent heavy-ion physics literature two other algorithms for
finding a surface in four dimensions have been described. In
Ref.~\cite{Hirano:2012kj} the surface elements are taken to be the
common faces of neighbouring volume elements when the isovalue is
reached between them. This algorithm is simple and robust, but it
requires very small grid spacing to avoid numerical artefacts (see
discussion in Refs.~\cite{Schenke:2010nt,Pang:2012he}).  ``Very
small'' in this context means that the volume element for the surface
extraction should be the same than the volume element for the fluid
evolution, whereas the more sophisticated algorithms discussed here
allow significantly larger volume elements for surface evaluation than
for fluid evolution. For example, in Ref.~\cite{Kolb:2000} the results
were calculated using timestep $\Delta\tau = 0.04$ fm and grid spacing
$\Delta x = \Delta y = 0.1$ fm, whereas the surface elements were
evaluated in cubes 10 times larger in time, and 2 to 4 times larger in
both spatial directions (depending on the impact parameter). In the
following, however, we play it safe and use the same grid spacing and
timestep for both the hydrodynamical evolution and surface extraction
to calculate the results shown in Sec.~\ref{sec:results}.

Another algorithm was described in Ref.~\cite{Pang:2012he} and called
a projection method. To our understanding this method insists that
there is only one surface element within each volume element. As
discussed above this may lead to double counting some parts of the
surface.

\section{Negative contributions of Cooper-Frye}
  \label{sec:negative}

One of the conceptual problems in using the Cooper-Frye procedure
(Eq.~(\ref{eq:CF})) to describe freeze-out in hydrodynamical models is
that if the surface element of the freeze-out surface is
spacelike\footnote{We use the convention to describe a surface as
  time- or spacelike according to its normal
  vector. $\dif\sigma_\mu\dif\sigma^\mu > 0$ is timelike, and
  $\dif\sigma_\mu\dif\sigma^\mu < 0$ is spacelike.}, particles with
certain momenta count as a negative contribution to the distribution:
These particles move inwards, they are not emitted at the surface but
absorbed\footnote{In the context of hydrodynamical models, these
  negative contributions have been discussed, and possible cures
  proposed, in
  Refs.~\cite{Rischke:1998fq,Cheng:2010mm,bugaev,cs_fo,Bernard:1996ba}.}.
However, in a hybrid model, these negative contributions are only a
technical, not a conceptual problem. In an ideal case, the switch from
fluid to cascade is done in a spacetime region where both descriptions
give (approximately) equal solutions~\cite{Bass:2000ib}. Thus, in the
vicinity of the switching surface the particle distributions are close
to thermal, and correspond to the hydrodynamical solution on both
sides of the surface. Assuming that the system is dilute enough for
the kinetic theory decomposition of the fluid variables in terms of
particle distributions to be valid, the particle distributions
$f(p,x)$ are known on the surface, and there are particles moving
inwards through the surface as described by Cooper-Frye. If this is
not the case, the particlization takes place in a region where
transport and hydro do not give equal solutions, and the switch from
hydro to cascade should take place elsewhere.

Thus as an initial state for the cascade, we have to count the
particles passing through the switching surface from hydro to
cascade. Their distribution is easily obtained by augmenting the
distribution in Eq.~(\ref{eq:CF}) by a $\Theta$-function counting
particles going outward:
\begin{equation}
  E\frac{\dif N(x)}{\dif p^3} 
        = \dif\sigma_\mu p^\mu f(x,p)\,\Theta(\dif\sigma_\mu p^\mu).
 \label{eq:theta}
\end{equation} 
We have to calculate the number of particles coming from hydro to
cascade using this distribution, but we must also have a drain term in
the cascade: All the particles crossing the particlization surface
from cascade to hydro must be removed from the cascade. Otherwise the
conservation laws are not obeyed. It is not possible to remove
particles from cascade at the rate described by Cooper-Frye procedure,
but we can calculate the distribution of particles which were removed
from cascade when they entered the fluid dynamical region. This
provides a consistency check: At the end of the evolution we can
compare whether the phase-space distribution of particles removed from
cascade is (approximately) equal to the negative contributions given
by Cooper-Frye. If they are, the calculation was consistent, if not,
then the switch from hydro to cascade took place in a wrong place, we
have to change the switching criterion, and redo the calculation.  Of
course this argument assumes that such a spacetime region exists where
transport and (dissipative) fluid dynamics give equal spacetime
evolution. If this is not the case the whole raison d'\^etre of hybrid
models is questionable, since the switching becomes as arbitrary as
the freeze-out in hydrodynamical models.

Unfortunately the treatment described above is technically
challenging. First, it is numerically expensive to evaluate the number
of particles emitted when the distribution is modified by the
$\Theta$-function (Eq.~(\ref{eq:theta})). Second, we are using UrQMD
to describe the hadronic transport, and the present version of UrQMD
does not allow tracing the propagation of individual particles in such
a way that whenever a particle crosses the boundary between cascade
and fluid dynamics, it would be removed from cascade. Such a feature
would require extensive rewriting of the code, and is therefore
postponed to later. Furthermore, we are using ideal hydro to describe
the fluid dynamical part, and we do not expect that there exists a
region where UrQMD would lead to ideal fluid dynamical
behaviour\footnote{For the difficulties in approaching the ideal fluid
  limit using transport, see \eg Ref.~\cite{Molnar:2004yh}.}.

Since the correct treatment is at the time of this writing beyond our
reach, we have to settle with an approximative solution like in
Ref.~\cite{Hirano:2012kj}: First, we ignore all the surface elements
where the flow is directed inwards, \ie where $\dif\sigma_\mu u^\mu <
0$, since the net number of particles passing through such a surface
element is negative. There are more particles moving inwards than
outwards. Second, on the surface elements where the flow is directed
outwards, $\dif\sigma_\mu u^\mu > 0$, we use the net number of
particles passing the surface, $\dif\sigma_\mu j^\mu$, as the number
of particles emitted, but we use the thermal distribution with
$\Theta$-function, Eq.~(\ref{eq:theta}) as their momentum
distribution. This leads to a small violation of conservation
laws. The emitted particles either have too large energy, or we
generate too few of them, depending on the constraints we impose on
the sampling procedure.

There are approaches in the literature where the problem of negative
contributions is solved differently. Instead of sampling the
phase-space distributions like we do here, in
VISHNU~\cite{Song:2010mg,Song:2010aq} and in
Ref.~\cite{Holopainen:2010gz}, the momentum distribution of each
particle species is calculated by integrating over the particlization
surface, Eq.~(\ref{eq:CF}), like in the conventional calculation of
particle spectra in hydrodynamical models~\cite{Huichao,Hannu}. The
number of particles of species $i$ per unit rapidity, $\dif N_i/\dif
y$, in these boost-invariant distributions is calculated, and the
sampling is constrained to produce either $10\dif N_i/\dif y$
particles in a wide rapidity interval of $-5 < y < 5$~\cite{Huichao},
or $\dif N_i/\dif y$ particles in an interval $-0.5 < y
<0.5$~\cite{Holopainen:2010gz}. These yields are not integers,
however. In VISHNU the yield, $10\dif N_i/\dif y$, is rounded down to
the nearest integer~\cite{Huichao}, whereas in
ref.~\cite{Holopainen:2010gz} the decimal part is taken as a
probability whether to create one additional particle over the rounded
down yield~\cite{Hannu}. At this stage the sampled particles carry no
information of their position. In Ref.~\cite{Holopainen:2010gz} there
are no rescatterings after particlization, and this lack of
information is not a problem, whereas in VISHNU each particle is given
a position according to the probability given by the Cooper-Frye
distribution, neglecting the part where the distribution would be
negative, Eq.~\ref{eq:theta}. Thus this approach reproduces the
Cooper-Frye momentum distribution, but it skews the position
distribution of the particles reducing the fraction of particles
emitted from timelike surface elements.  How this affects the final
distribution of particles after rescatterings is unknown. Another
disadvantage of this approach is that since it requires evaluating the
momentum distributions of all particle species by integrating over the
particlization surface, it is slow.

As mentioned in the beginning of this section we want to switch from
fluid to cascade in a region where both approaches give similar
results, and that there is no physical change in the region where the
switch takes place. This means that the solutions to the negative
contributions at freeze-out proposed in
Refs.~\cite{Cheng:2010mm,bugaev,cs_fo} are not applicable in our
case. In those approaches a non-thermal freeze-out distribution which
contains no particles moving inwards is postulated. But since we
assume that the distributions are the same on both sides of the
switching surface, such a distribution on the particle side would be a
contradiction. Furthermore it is perfectly possible that the hadrons
in the cascade scatter back to the spacetime region described by fluid
dynamics, and assuming that there are no particles entering that
region is an oversimplification.

\section{Sampling the distributions}
  \label{sec:sampling}

Any hadron cascade requires an particle ensemble with well defined
particle species, momenta, and positions as an initial state. To
generate such an ensembles we Monte-Carlo sample the particle
distributions on the particlization surface. For this purpose
we employ two different sampling algorithms. The first one that we
call 'allcells' sampling, contains a loop over all hypersurface
elements and in each of the elements, \ie cells, the particles are
sampled according to the steps explained below. In this case the
quantum numbers like energy, net baryon number, net strangeness and
electric charge are only conserved on the average over many sampled
events. The other algorithm that is dubbed 'mode' sampling introduces
a way to conserve all these quantities event-by-event, as we will
demonstrate in Section \ref{sec:cons_quantum}.  The whole sampling
algorithm is based on previous work published in
\cite{Petersen:2008dd}, but now applied without the assumption of an
isochronous transition and with slight improvements.

In general fluctuations in conserved quantum numbers occur in
event-by-event studies when only part of the system is described. For
example, the spectators are usually not included in the hydrodynamical
description of the system. As well fluctuations occur in a rapidity
interval which does not contain all the emitted particles. But, once
an initial state of the system is defined, and the whole subsequent
evolution is considered, the conserved quantum numbers must be
conserved. The straightforward 'allcells' sampling does not do it, and
therefore we have developed the more complicated 'mode' sampling
approach.

The first step in sampling hadrons on the hypersurface is to decide in
which cells a particle has been produced. For example, in a central
Au+Au collision at the highest RHIC energy there are roughly $10^7$
hypersurface elements, but only $\sim$ 10,000 particles
produced. Evaluating first, if there is a particle produced in a cell,
before doing any of the other steps, results in a speed-up of the
calculation, which is essential for the application to event-by-event
calculations.

The number of particles of each hadron species produced in one cell is
calculated according to the following formula (only the surface
elements with $u^\mu \dif\sigma_\mu>0$ are considered) :
\begin{equation}
N_i = j^\mu \dif\sigma_\mu = n_i  u^\mu  \dif\sigma_\mu\,,
\label{eqn_nop}
\end{equation}
where $n_i$ is the particle density in the local rest frame and $i$ is
the index of the particle species. It is important to take into
account all 150 particle species and their antiparticles implemented
in UrQMD to match the equation of state on the particlization
surface. Assuming a Boltzmann distribution the integral over momentum
space for the particle number density in the local rest frame can be
evaluated analytically and the result is
\begin{equation}
n_i=\frac{4\pi g_i m_i^2 T}{(2\pi)^3}e^{\frac{\mu}{T}}K_2\left(\frac{m_i}{T}\right)\,,
\end{equation}
where $g_i$ is the degeneracy factor for the respective particle
species. All the information about the particle properties is read in
directly from the UrQMD tables to avoid any mismatch. The chemical
potential includes the baryo-chemical potential and the strangeness
chemical potential in the following way
\begin{equation}
\mu=B\cdot \mu_B+S\cdot \mu_S\,,
\end{equation}  
where S is the quantum number for strangeness and B is the baryon
number. For pions the Bose distribution has to be taken into account
because their mass is on the order of the temperature of the system.
Expanding the distribution in series and integrating leads to
\begin{equation}
n_\pi = \frac{g_\pi m_\pi^2T}{(2\pi)^2}\sum_{k=1}^\infty \frac{1}{k}K_2\left(\frac{k
m_\pi}{T}\right)\,,
\end{equation} 
where we stop the summation after 10 summands, when it has
sufficiently converged.
 
By summing up all the contributions from different species the total
number of particles in this cell $N=\sum N_i$ is known. As long as 
$N< 0.01$ one can interpret this number directly as a sampling
probability and randomly decide, if a particle is produced or not. If
$N$ is larger, one needs to sample a Poisson distribution with $N$ as
the mean value to decide the actual number of particles produced in
this surface element. It turns out that the cases where 2 or more
particles per cell are produced are rare.  If there is a particle
produced the species can be decided by sampling according to their
probabilities $N_i/N$. Isospin is assigned randomly consistent with
the isospin symmetry of a system in chemical equilibrium.

The four momenta of the particles are sampled according to the local
Cooper-Frye distribution (only the parts where 
$f(x,p) p^\mu \dif\sigma_\mu >0$ are considered)
\begin{equation}
 \frac{\dif N(x)}{\dif^3p} = \frac{1}{E} f(x,p) p^\mu \dif\sigma_\mu\,,
\end{equation}
where f(x,p) are the boosted Fermi or Bose distributions corresponding
to the respective particle species including again the chemical
potentials for baryon number and strangeness.  Finding the maximum of
the distribution and then applying the rejection method is crucial
since the distribution functions in momentum space are highly peaked
and dependent on the masses and thermodynamic parameters. Currently,
an approximative maximum of
\begin{equation}
\label{eqn:mom_dist}
\frac{1}{E} p^\mu \dif\sigma_\mu f(x,p)
\end{equation}
is determined by a coarse loop over the three-dimensional momentum
space. This approximate maximum is multiplied by 1.2 to make sure it
is definitely larger than the function to be sampled. 

Then, a momentum vector $\mathbf{p}$ is chosen randomly, and an
additional random number between zero and the above mentioned maximum
of the distribution is generated. If this random number is smaller
than the value of the distribution at this momentum, the momentum is
accepted. If not, another momentum and another random number are
generated, and the process is repeated until an acceptable momentum is
found.

As described in Sec.~\ref{sec:negative} we neglect negative parts of
the distribution functions which slightly alters the resulting
rapidity and transverse momentum spectra. It is important to sample
the momenta according to the boosted distribution function instead of
sampling the equilibrium distribution in the local rest frame and then
boost the momentum four-vector back to the computational frame. The
second procedure leads to a violation of energy conservation since one
does not reproduce the whole tensor $T^{\mu\nu}$ in the computational
frame correctly \cite{Cooper:1974mv}.

Imposing strict conservation laws on grand canonical distributions on
event-by-event basis is not wholly consistent, and we minimise any
bias it creates by the following procedure: To conserve energy, net
baryon number, net strangeness and electric charge in all events we do
seven subsequent random loops called modes over the hypersurface.
During the first mode, cells are randomly chosen and particles are
sampled until the total energy is conserved. From this set of
particles only the ones containing a $\bar{s}$ quark are kept, and the
rest are discarded. In the second mode we produce particles in a
similar way, until we've produced as much anti-strangeness as the
first mode produced strangeness.  We keep the anti-strange particles,
and discard the rest. In modes three and four we repeat the same
procedure keeping the non-strange baryons and non-strange antibaryons,
respectively, but requiring that the net baryon number of these
particles is the net baryon number of the system. Modes five and six
take care of the conservation of electric charge by keeping the
negative and positive non-strange mesons, respectively, and finally
mode seven takes care of the energy conservation by sampling neutral
non-strange mesons until the energy of the particles corresponds to
the energy of the fluid.

\section{Results}
  \label{sec:results}
\subsection{Description of the Hybrid Model}
\label{sec:hybrid}

The hypersurface finding and sampling algorithms described in
Sections~\ref{sec:surface} and~\ref{sec:sampling} have been
implemented in the hybrid model of Ref.~\cite{Petersen:2008dd}. We
show here some preliminary results to demonstrate some general trends
and overall behaviour of the improved model. Note that we have not
tried to fine-tune any parameters to describe experimental data at
this point.

Smooth initial conditions for the hydrodynamic evolution have been
generated by averaging over 100 UrQMD events that are run up to the
starting time of $t_{\rm start}=2.83$ fm for Pb+Pb collisions at
$E_{\rm lab}=160A$ GeV (SPS) \cite{Petersen:2010md} and 
$t_{\rm start}=0.5$ fm for Au+Au collisions at $E_{\rm cm}=200A$ GeV
\cite{Petersen:2010zt} (RHIC). 
 
On the constant time surface $t=t_{\rm start}$ the energy, momentum
and net baryon number densities are determined by representing each
particle in the UrQMD event with a three-dimensional, longitudinally
Lorentz-contracted, Gaussian density distribution of thermalised
matter, which has the same energy, momentum and baryon number than the
particle it represents. Summing up the distributions representing the
particles, we obtain a distribution of matter in thermal equilibrium,
which has the same energy, momentum and baryon number than the UrQMD
event in question. For the RHIC initial state we consider only
particles within the rapidity interval $-2<y<2$, whereas we include
all particles that have interacted at least once when evaluating the
initial state for SPS. The mean value of these distributions from 100
events leads to a smooth profile but still includes finite initial
velocities in all three directions~\cite{Petersen:1900zz}.

The (3+1)D ideal hydrodynamic evolution is solved in Cartesian
coordinates using the SHASTA
algorithm~\cite{Rischke:1995ir,Rischke:1995mt}. The equation of state
is calculated within a chiral model coupled to the Polyakov loop to
include the deconfinement transition that reproduces nuclear ground
state properties and lattice QCD data at zero net-baryon chemical
potential \cite{Steinheimer:2010ib}. This equation of state is also
applicable at finite net baryon densities, which allows us to
calculate heavy ion collisions at SPS energies to explore the beam
energy dependence of our findings. In the hadronic phase this equation
of state is equivalent to the effective equation of state of UrQMD,
and has the same degrees of freedom than UrQMD. The conservation laws
are thus automatically fulfilled on particlization surface when we use
Cooper-Frye prescription to calculate the particle distributions.

The hypersurface finder 'cornelius' has been implemented in the
hydrodynamic code in such a way that the hypersurface for any
transition criterion is evaluated while the hydrodynamic evolution is
calculated. Then we apply the two sampling algorithms and further
calculate resonance decays using UrQMD. To obtain final results that
can be compared to experimental data the hadronic transport approach
is used in addition to calculate the rescatterings.

\subsection{Structure of the Hypersurface}
 \label{sec:structure}

The equations of motion of relativistic hydrodynamics are nothing more
than an application of conservation laws for energy, momentum and
charge(s). Thus one of the first checks on the accuracy of the
numerical solutions of hydrodynamics is to confirm the validity of the
conservation laws.

\begin{table}
  \caption{The conservation of energy (E) and net baryon number (B) in
    Au+Au collisions at $E_{\rm cm}=200A$ GeV. The values in the final
    state are split into two parts: \emph{pos.} is flow
    through elements where the energy or baryon flow is directed outwards,
    $\dif\sigma_\mu T^{\mu 0} >0$ or $\dif\sigma_\mu n_B u^\mu >0$, respectively,
    whereas \emph{neg.} is flow through elements where energy or baryon flow
    is directed inwards, $\dif\sigma_\mu T^{\mu 0} <0$ or
    $\dif\sigma_\mu n_B u^\mu<0$, respectively, and thus counts as negative.
    Note that these are not the negative contributions of Cooper-Frye, see
    the text. The upper two rows are for central ($b<3.4$ fm) and the lower
    two rows for mid-central ($b=7$ fm) collisions.}
\label{tab_cons_rhic}
\begin{tabular}{|l|c|c|c|c|c|c|}
\hline
 & \multicolumn{3}{c|}{E [GeV]} & \multicolumn{3}{c|}{B}\\
\hline
 & total & \emph{pos.} & \emph{neg.} & total & \emph{pos.} & \emph{neg.}\\
\hline
\hline
initial& 5431 &      &      & 93.23 &       &       \\
final  & 5430 & 5861 & -431 & 92.74 & 97.74 & -5.00 \\
\hline
\hline
initial& 2327 &      &      & 35.84 &       &       \\
final  & 2336 & 2455 & -119 & 35.80 & 37.10 & -1.30 \\
\hline
\end{tabular}
\end{table}

\begin{table}
  \caption{The conservation of energy (E) and net baryon number (B) in
    Pb+Pb collisions at $E_{\rm lab}=160A$ GeV. The layout is the same
    as in Table~\ref{tab_cons_rhic}.}
\label{tab_cons_sps}
\begin{tabular}{|l|c|c|c|c|c|c|}
\hline
 & \multicolumn{3}{c|}{E [GeV]} & \multicolumn{3}{c|}{B}\\
\hline
 & total & \emph{pos.} & \emph{neg.} & total & \emph{pos.} & \emph{neg.}\\
\hline
\hline
initial& 3117 &      &       & 347.6  &       &      \\
final  & 3120 & 3208 & -88.5 & 345.9  & 355.8 & -9.9 \\
\hline
\hline
initial& 1528 &      &       & 170.23 &        &       \\
final  & 1528 & 1570 & -41.5 & 169.22 & 174.26 & -5.04 \\
\hline
\end{tabular}
\end{table}

We show the total energy (E) and net baryon number (B) in collisions
at two different centralities at RHIC and SPS in tables
\ref{tab_cons_rhic} and \ref{tab_cons_sps}, respectively. The initial
value is evaluated in the beginning of the hydrodynamical evolution by
summing over all fluid cells within the particlization surface. With
the switching criterion we use here, $\epsilon = 2\epsilon_0$, where
$\epsilon_0$ refers to the nuclear ground state energy density of
$~146$ MeV/fm$^3$, this means summing over the cells with 
$\epsilon > 2\epsilon_0$. Since the final state we are interested in
is the isosurface with $\epsilon = 2\epsilon_0$, we evaluate the final
state energy and net baryon number as energy and baryon number flows
through this surface:
\begin{equation}
 E=\int_\sigma T^{0\mu} \dif\sigma_\mu \quad \mbox{and} \quad 
 B=\int_\sigma n_B u^\mu \dif\sigma_\mu, 
\end{equation}
where the energy momentum tensor has been evaluated as
$T^{\mu\nu}=(\epsilon +P) u^\mu u^\nu-g^{\mu\nu} P$. As seen in the
tables both the energy and baryon number are nicely conserved with
better than 0.6\% accuracy in all cases.

However, a closer look at the properties of the particlization surface
reveals that there are surface elements where the flow is directed
inwards, $\dif\sigma_\mu u^\mu < 0$. As described in
Sec.~\ref{sec:sampling}, such elements cannot be used as a source for
sampled particles, and thus we check how large uncertainty these
elements cause to the total energy and baryon number.  In tables
\ref{tab_cons_rhic} and \ref{tab_cons_sps}, the flows through the
surface are also separated in two parts: \emph{pos.} depicts the
energy and baryon flows through surface elements where 
$\dif\sigma_\mu T^{\mu 0} > 0$ for energy and 
$\dif\sigma_\mu n_B u^\mu > 0$ for baryon flow, whereas \emph{neg.} is
the flow through elements where $\dif\sigma_\mu T^{\mu 0} < 0$ for
energy and $\dif\sigma_\mu n_B u^\mu < 0$ for baryons. As seen the
surface elements where flow is directed inwards cause a 3--8\%
uncertainty in the total energy, and a 3--5\% uncertainty in the total
baryon number of the system. Note that because of the pressure term in
$T^{00}$ it is possible that we have an element where energy flow is
directed inwards but baryon flow outwards, or vice versa.

\begin{figure}
\resizebox{0.5\textwidth}{!}{\includegraphics{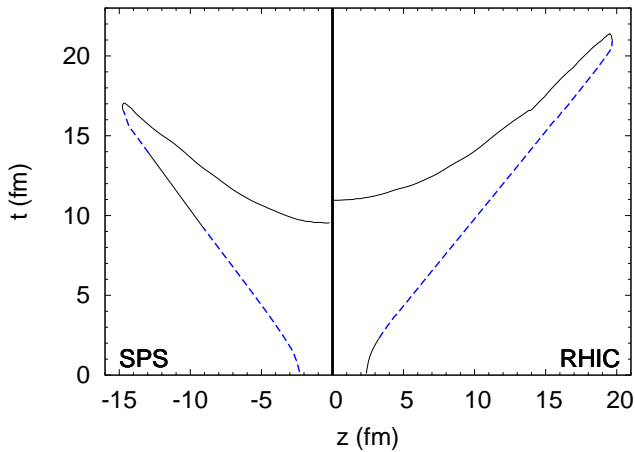}}
\caption{The position of the $\epsilon = 2\epsilon_0$ isosurface in
  $zt$-plane at $x=y=0$. The left side of the plot (SPS) depicts
  the surface in central Pb+Pb collisions at $E_{\rm lab}=160A$ GeV
  and the right side (RHIC) in central Au+Au collisions at
  $E_{\rm cm}=200A$ GeV. The dashed lines depict regions where the
  flow may be directed inwards on some surface elements.}
\label{fig:zt-surface}
\end{figure}

To understand this phenomenon we located the surface elements with
$\dif\sigma_\mu u^\mu < 0$ and found out that they are almost entirely
located at the edges of the system where the fireball expands
longitudinally, see Fig.~\ref{fig:zt-surface}, and there are none at
midrapidity. Furthermore, the regions depicted by dashed lines in
Fig.~\ref{fig:zt-surface} are not regions where the flow is directed
inwards, but they are regions where surface elements with inwards
directed flow randomly appear among elements where the flow is
directed outwards. These are the regions where the longitudinal
pressure gradient is still large, the fluid flow is very large, $v_z
\sim 0.9$--0.95, and the isosurface moves outwards with a comparable
speed as well. It is known that this is a difficult terrain for SHASTA
and for any algorithm to solve accurately~\cite{Rischke:1995ir}. Since
the flow velocity is almost aligned with the surface, it does not
require a large numerical error to flip an outwards flow to
inwards. Thus it is possible that these elements with flow directed
inwards are just numerical artefacts. However, it is worth
remembering that in event-by-event calculations the initial states are
highly irregular, and occasionally flow can be directed inwards for
physical reasons.

\begin{figure}
\resizebox{0.5\textwidth}{!}{\includegraphics{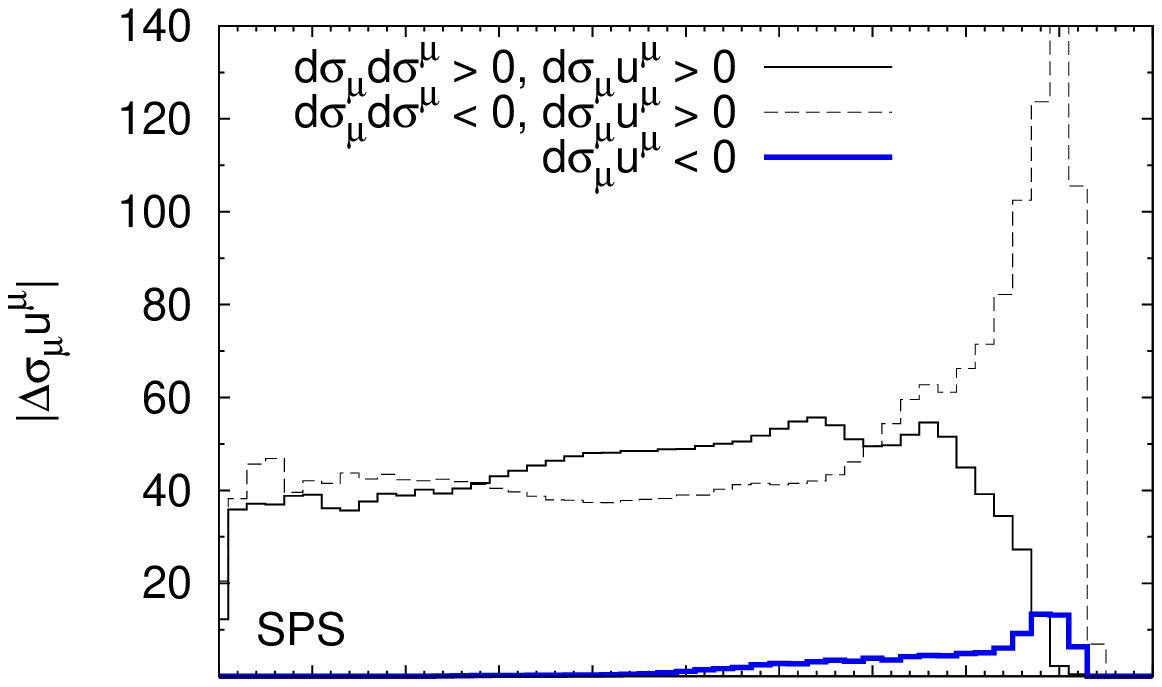}}
\resizebox{0.5\textwidth}{!}{\includegraphics{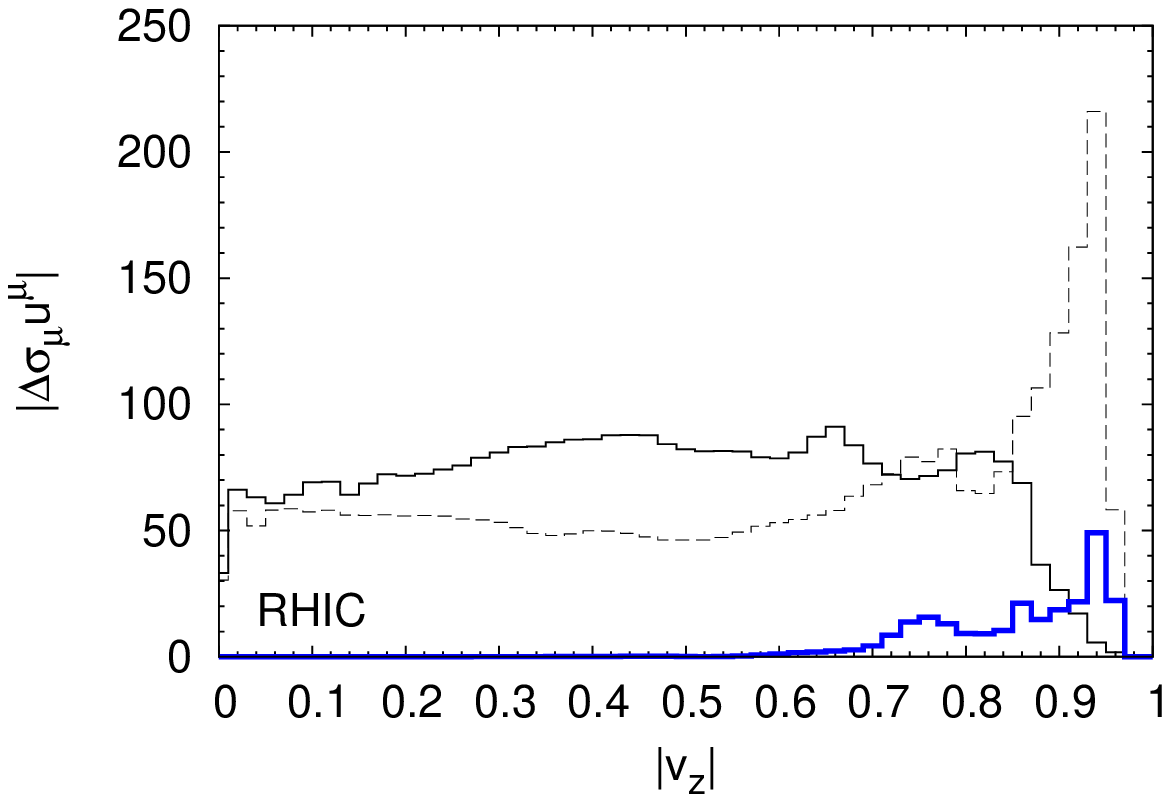}}
\caption{The magnitude of the fluid flow through different parts of
  the particlization surface at different longitudinal flow velocities
  in central Pb+Pb collisions at $E_{\rm lab}=160A$ GeV (SPS) and in
  central Au+Au collisions at $E_{\rm cm}=200A$ GeV (RHIC), see the
  text.}
\label{fig:flow}
\end{figure}

Since the common feature of the elements with inwards directed flow is
the large longitudinal flow velocity, we further characterise the
surface by it. We bin the surface elements according to the magnitude
of the longitudinal flow velocity $|v_z|$, and integrate over the
elements in each bin to obtain the magnitude of fluid flow through
surface, $|\dif\sigma_\mu u^\mu|$, at different velocities
$|v_z|$. The results are shown in Fig.~\ref{fig:flow}. The
contribution from elements with flow directed inwards is depicted by
the thick curve, and as claimed it is concentrated at large
velocities. Furthermore it can be seen that the fluid flow inwards is
small compared to the fluid flow outwards, which is depicted by the
thin solid and dashed histograms in Fig.~\ref{fig:flow}. The 5--8\%
uncertainties in energy and baryon number listed in tables
\ref{tab_cons_rhic} and \ref{tab_cons_sps} are thus concentrated at
large longitudinal flow velocities, and can be expected to have only a
tiny contribution to observables at midrapidity. The reason why there
is a difference between fluid flow $\dif\sigma_\mu u^\mu$ and energy
and baryon number flows, $\dif\sigma_\mu T^{\mu 0}$ and
$\dif\sigma_\mu j^\mu$, respectively, is that the former has an
additional gamma factor and a pressure term, whereas the latter is also
sensitive to the baryon chemical potential $\mu_B$ which is not
uniform on the surface.

\subsection{Negative contributions}
 \label{sec:neg}

It is worth remembering that the negative contributions to energy and
baryon number discussed above in Sec.~\ref{sec:structure} are not the
same thing than the negative contributions of Cooper-Frye discussed in
the literature (see Sec.~\ref{sec:negative}). The former are caused by
the collective flow being directed inwards, whereas the negative
contributions of Cooper-Frye are caused by individual particles moving
inwards through the particlization surface. As discussed in
Sec.~\ref{sec:sampling}, the sampling routine cannot take into account
particles moving inwards nor anything emitted on surfaces where flow
is directed inwards. Therefore we evaluated these negative
contributions to see how much omitting them distorts the spectra.

We evaluated the spectra of thermal pions, kaons and protons on the
particlization surface using the conventional Cooper-Frye procedure,
Eq.~(\ref{eq:CF}), \ie we integrated over the surface to obtain
distributions, but we divided the contribution into four parts
according to whether the flow was directed in- or outwards
($\dif\sigma_\mu u^\mu < 0$ or $> 0$) and whether the particles with the
momentum in question were heading in- or outwards 
($\dif\sigma_\mu p^\mu < 0$ or $> 0$). It turned out that in general
the negative contributions to pion distribution are largest, so we
concentrate on them. Some of the actual spectra are shown in
Sec.~\ref{sec:tests} where we compare the sampled distributions to the
integrated ones. Here we emphasise the size of the negative
contributions by showing their relative contribution to the total
spectrum.

\begin{figure}
 \begin{center}
  \resizebox{0.45\textwidth}{!}{\includegraphics{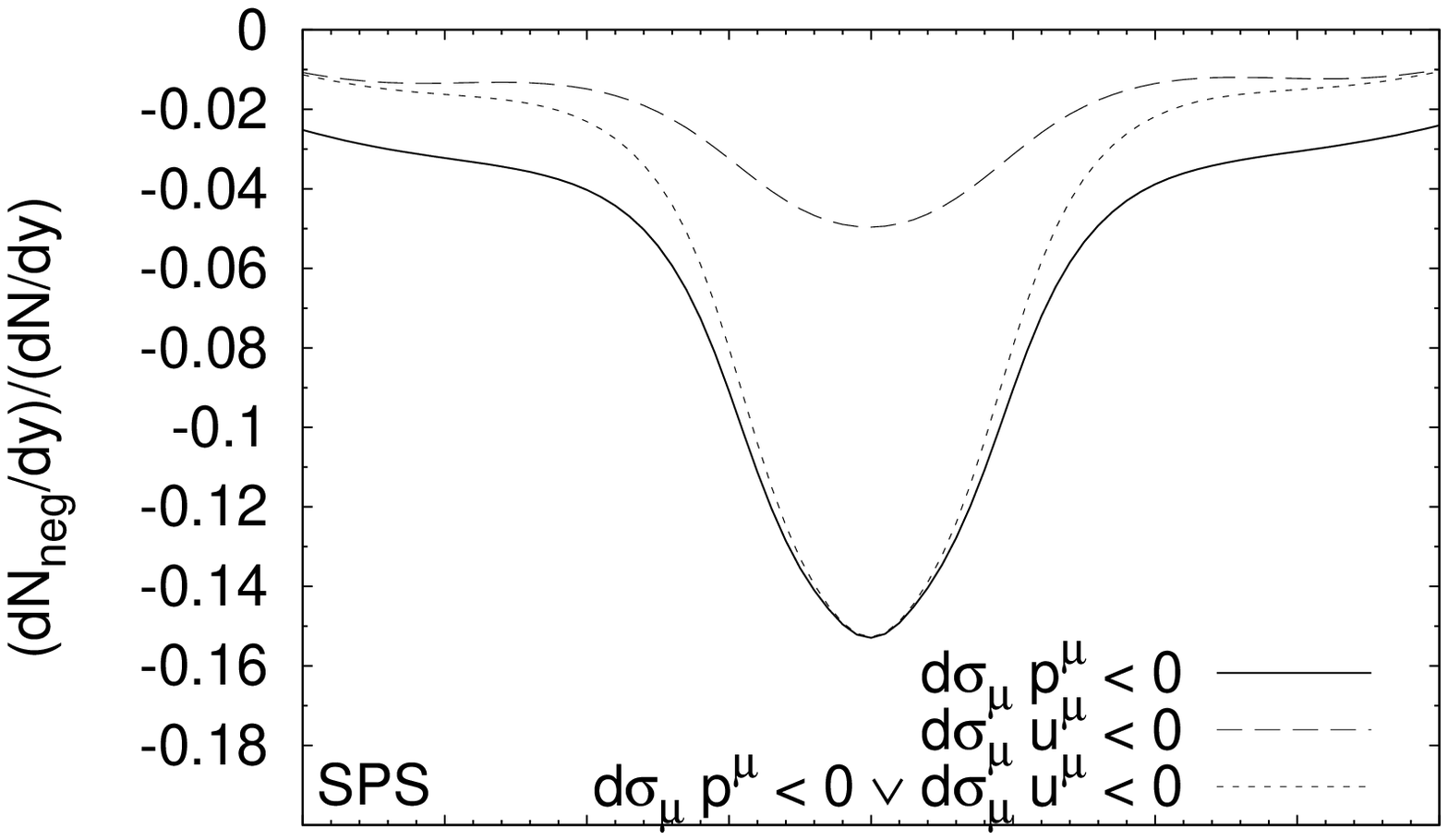}}
  \resizebox{0.45\textwidth}{!}{\includegraphics{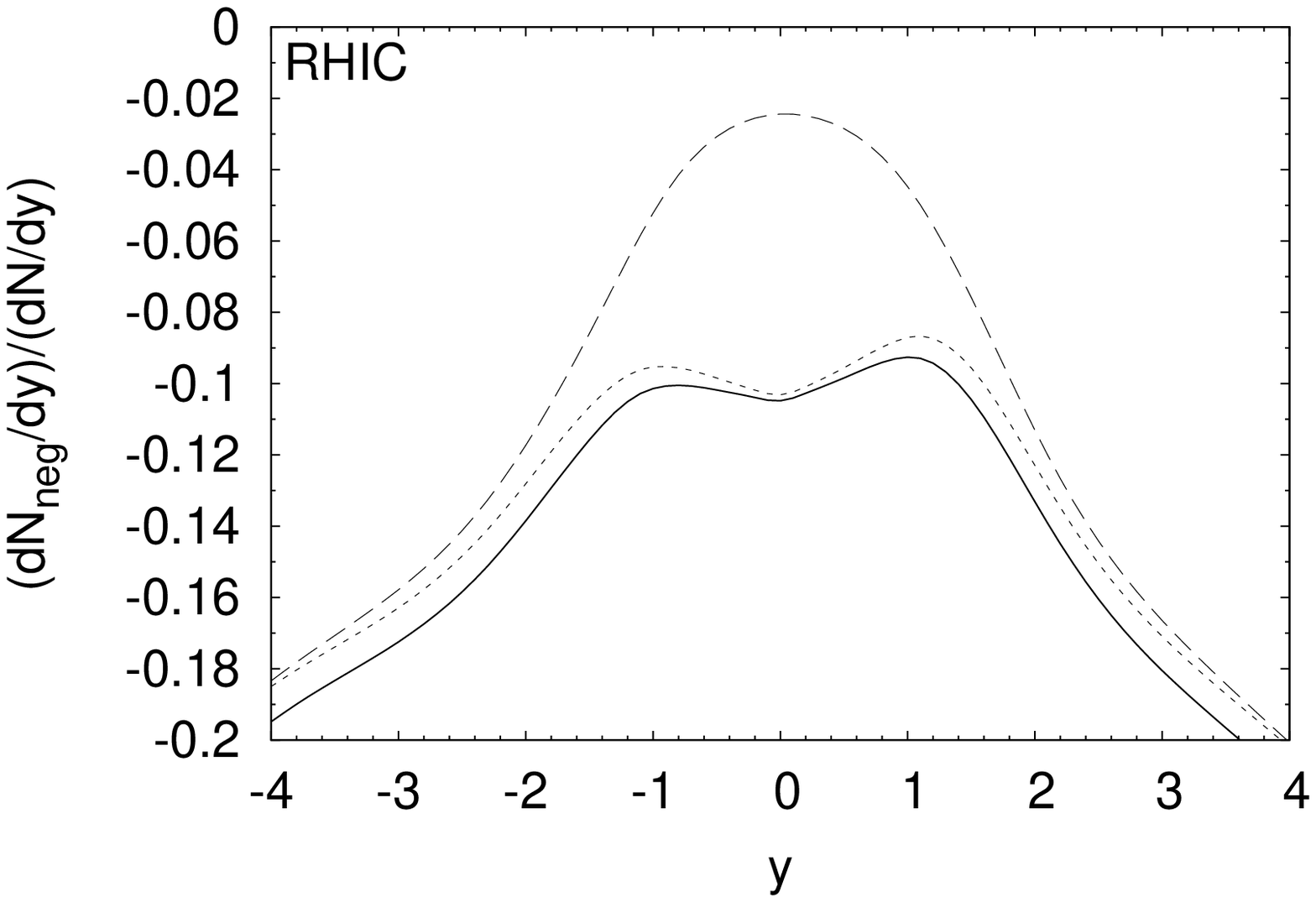}}
 \end{center}
 \caption{The ratio of the negative contribution to the total thermal
   pion rapidity distribution in central Pb+Pb collisions at 
   $E_{\rm lab}=160A$ GeV (SPS) and in central Au+Au collisions at
   $E_{\rm cm}=200A$ GeV (RHIC). The curves correspond to different
   kinds of contributions, see the text.}
\label{fig:negrap}
\end{figure}

In Fig.~\ref{fig:negrap} different negative contributions to the
thermal pion rapidity spectrum are shown for collisions at RHIC and
SPS. To check how much uncertainty the contribution from elements with
flow directed inwards cause, we have calculated the contribution from
them (dashed line, $\dif\sigma_\mu u^\mu > 0$) even if contribution
from them is not negative in the Cooper-Frye sense. It is seen that at
midrapidity contribution from them is ~2--5\%. A modest contribution
at midrapidity could be expected, since as shown in
Fig.~\ref{fig:flow}, the longitudinal flow velocity on these elements
is large. Interestingly the contribution from these elements at large
rapidities is very different at RHIC and SPS: The magnitude of the
relative contribution decreases with increasing rapidity at SPS, but
increases at RHIC. Again a result of different flow and emission
patterns shown in Fig.~\ref{fig:flow}.

We may take the surface and flow pattern as granted, and simply
evaluate the fraction of inward moving particles. This is depicted by
the solid line, $\dif\sigma_\mu p^\mu < 0$, in
Fig.~\ref{fig:negrap}. At midrapidity the contribution of such a
particles is surprisingly large, 10--15\% with a larger contribution
at SPS than at RHIC. That the negative contribution is larger at SPS
was already implied in Fig.~\ref{fig:flow} where the fluid flow
through the particlization surface was divided into a flow through
time- and spacelike surfaces. At SPS the relative fraction of the
fluid flow passing through spacelike surfaces is larger than at
RHIC. Since negative contributions appear only on spacelike surfaces,
the flow pattern at SPS provides a possibility for larger negative
contributions than at RHIC. Furthermore, if surfaces are similar, the
lower the flow velocity, the larger the negative contributions. At
SPS the flow velocity is lower than at RHIC, and thus it is rather
surprising that the difference in negative contributions is not larger
than what is shown in Fig.~\ref{fig:negrap}.

Finally, we want to check how large contribution our sampling routine
misses. Since it cannot sample the particles going inwards nor
particles emitted when the flow is directed inwards, sampling misses
both contributions. This is shown as the dotted line in
Fig.~\ref{fig:negrap}. At midrapidity the missed part is totally
dominated by the conventional negative Cooper-Frye contribution. Thus
it is safe to say that what comes to sampling particles at
midrapidity, the inwards directed flow is not a problem, but the
conventional negative contributions of Cooper-Frye can be.

\begin{figure}
 \begin{center}
  \resizebox{0.45\textwidth}{!}{\includegraphics{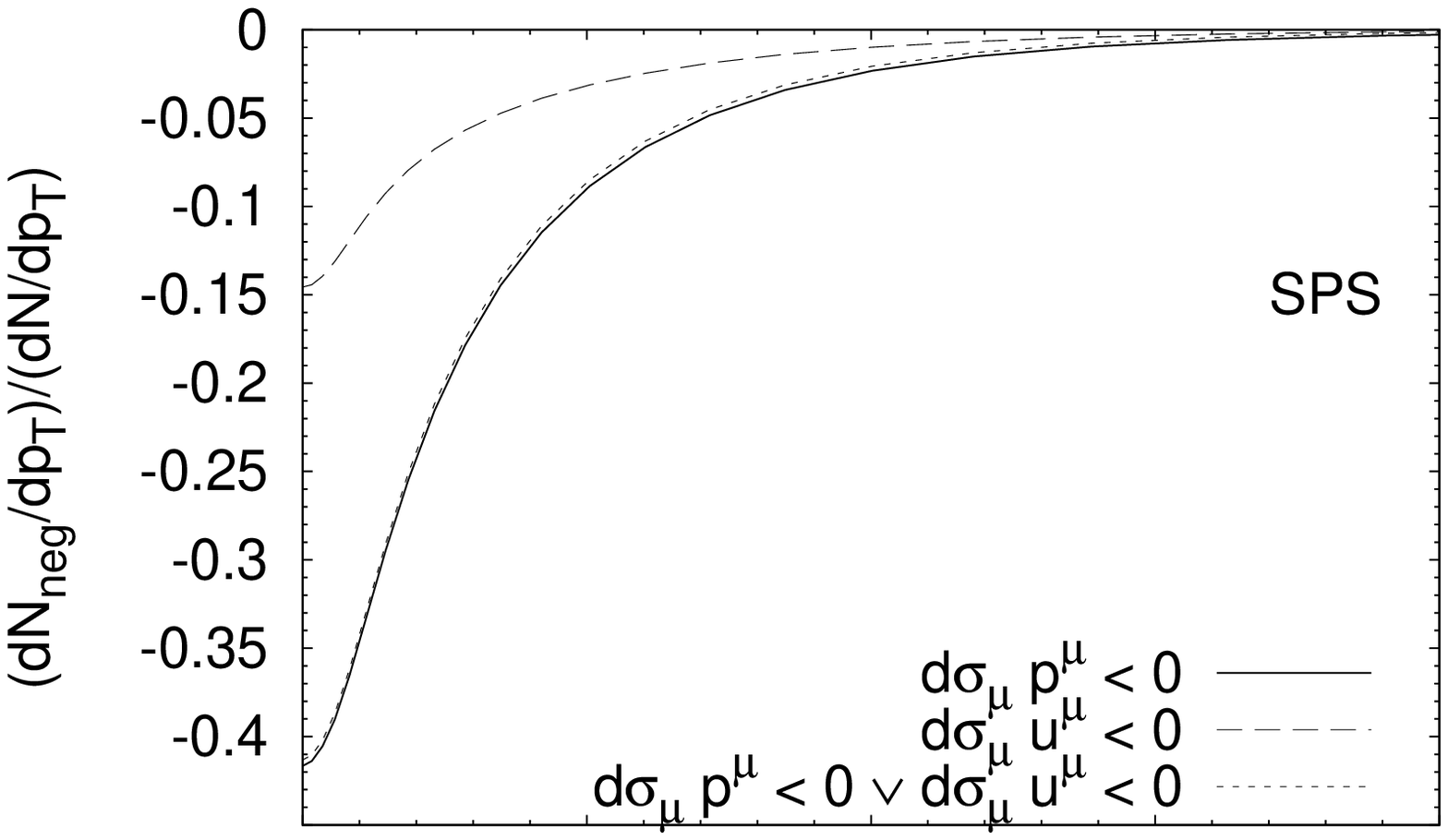}}
  \resizebox{0.45\textwidth}{!}{\includegraphics{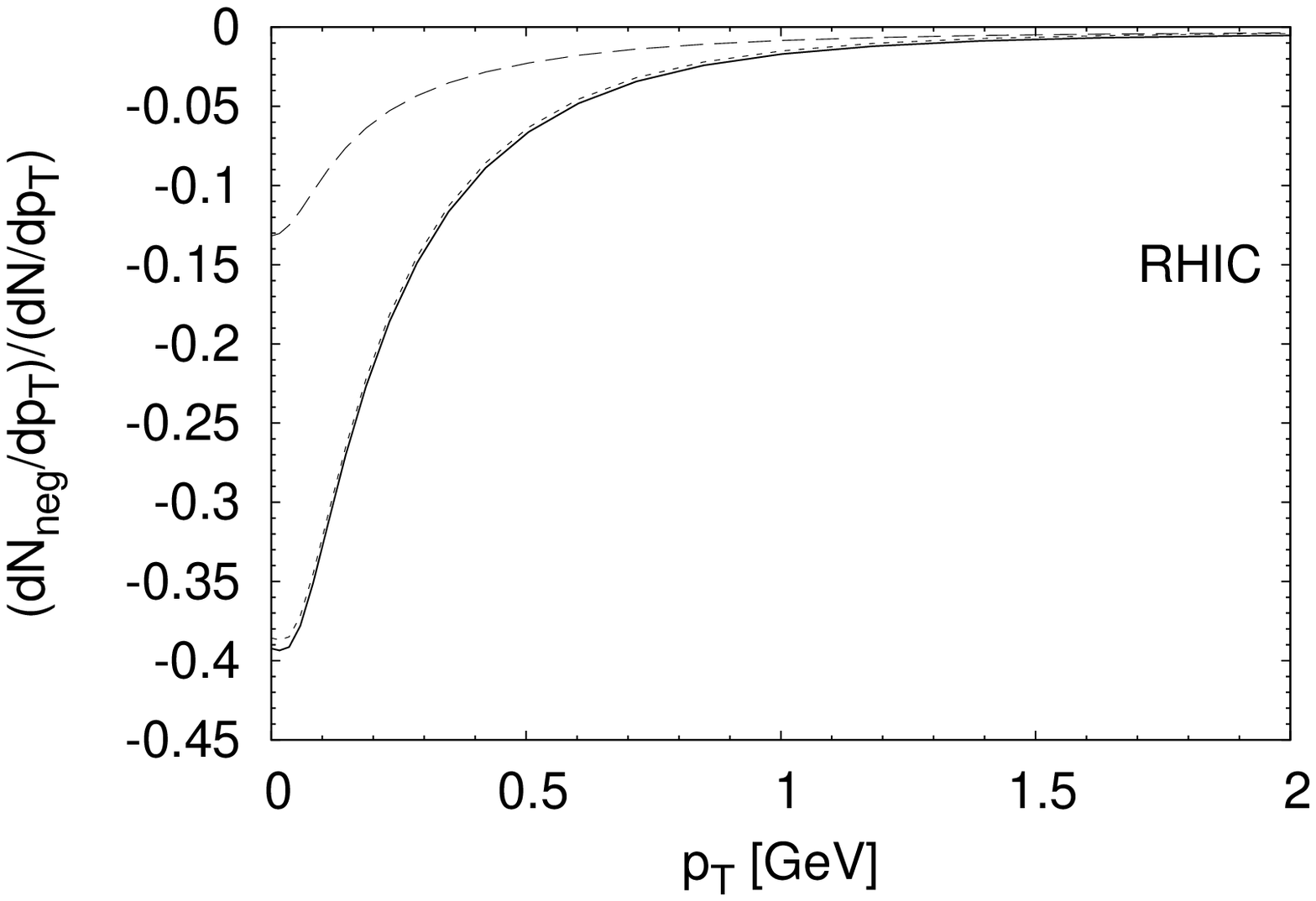}}
 \end{center}
 \caption{The ratio of the negative contribution to the total thermal
   pion $p_T$-distribution at midrapidity $-1<y<1$ in central Pb+Pb
   collisions at $E_{\rm lab}=160A$ GeV (SPS) and in central Au+Au
   collisions at $E_{\rm cm}=200A$ GeV (RHIC). The curves correspond
   to different kinds of contributions, see the text.}
\label{fig:negpt}
\end{figure}

To further study the effect of negative contributions, we show the
contributions to $p_T$ distribution of thermal pions at midrapidity,
$-1<y<1$, in Fig.~\ref{fig:negpt}. Again we see that the contribution
from the inwards directed flow is smaller than from inwards moving
particles. The latter can even reach ~40\% at low values of transverse
momentum!  Fortunately the size of the negative contribution decreases
rapidly with increasing transverse momentum, and already around 500
MeV they are much more acceptable 8--10\%. That the negative
contributions are largest at small $p_T$ is understandable since the
high $p_T$ particles are mostly emitted in regions where the flow
velocity is large and parallel to momentum, whereas the negative
contributions arise when the momentum is antiparallel to flow
velocity.

We also evaluated the $p_T$ averaged $v_2$ of thermal pions at
particlization in mid-central ($b=7$ fm) collisions at RHIC and
SPS. When $v_2$ was evaluated in a conventional fashion, we got $v_2 =
0.064$ and 0.066 at RHIC and SPS, respectively. When we removed the
negative contributions from the distribution, $v_2$ at RHIC and SPS
decreased to 0.058 and 0.06, respectively. If we also removed the
remaining positive contribution from elements with flow directed
inwards, further change on $v_2$ was on the level of $10^{-4}$. This
additional uncertainty of ~10\% should be kept in mind when discussing
the elusiveness of the QGP shear viscosity~\cite{Shen:2011zc}.

\subsection{Conservation of Quantum Numbers}
\label{sec:cons_quantum}

\begin{figure}
\resizebox{0.5\textwidth}{!}{
  \includegraphics{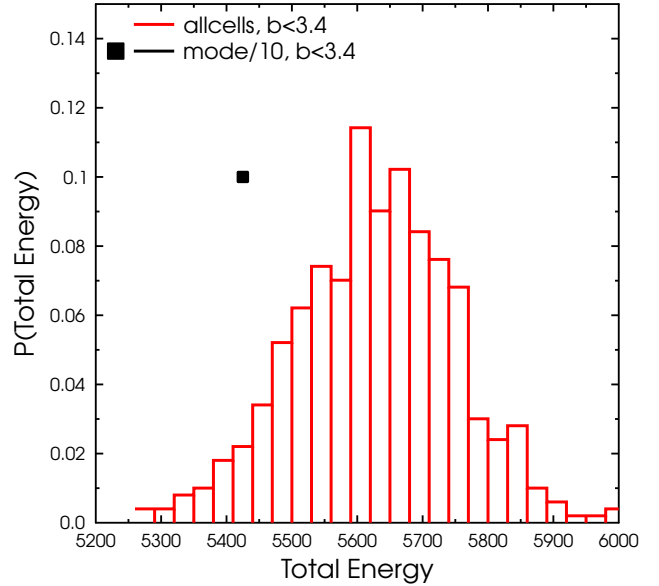}
}
\caption{Probability distribution for the total energy of an event in
  an ensemble of 500 events created using two different sampling
  algorithms in central Au+Au collisions at $E_{\rm cm}=200A$ GeV.}
\label{fig:energy_dist}       
\end{figure}

\begin{figure}
\resizebox{0.5\textwidth}{!}{
  \includegraphics{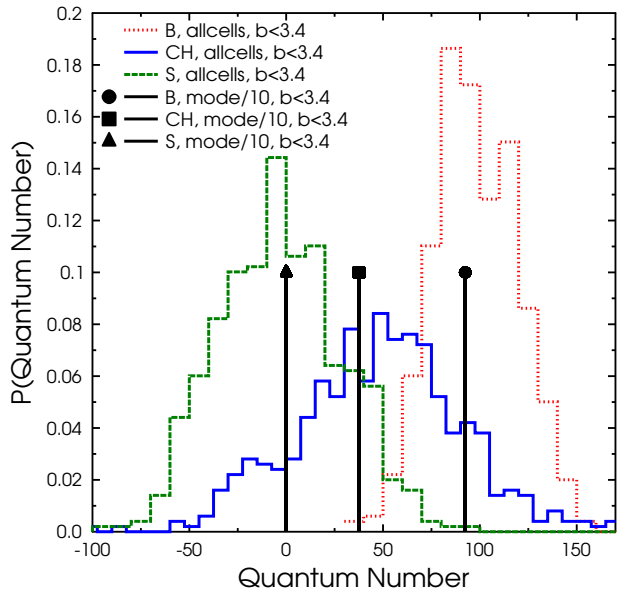}
}
\caption{Probability distribution for the total net baryon number (B),
  net strangeness (S) and net charge (CH) of an event in an ensemble
  of 500 events created using two different sampling algorithms in
  central Au+Au collisions at $E_{\rm cm}=200A$ GeV.}
\label{fig:quantum_dist}       
\end{figure}

In Figs. \ref{fig:energy_dist} and \ref{fig:quantum_dist} we show the
probability distributions of relevant quantum numbers using the two
sampling algorithms described in Section \ref{sec:sampling}. As
explained, the 'mode' sampling is constructed to obey conservation
laws event-by-event, whereas in the simpler sampling of all
hypersurface elements the conserved quantities fluctuate around the
values given by the positive contributions to the spectra. Note that
these values are not those given in Tables~\ref{tab_cons_rhic}
and~\ref{tab_cons_sps} as energy and baryon number flow through the
parts of surface where flow is directed outwards, but by the
distributions of particles coming out through these parts of the
surface, Eq.~(\ref{eq:theta}), after these distributions have been
scaled to yield the same number of particles than the local net flow
of particles through the surface element, $\dif\sigma_\mu nu^\mu$.
Therefore, we expect to see a higher mean value in the energy and net
baryon number in the allcells sampling compared to the mode setup. One
of the reasons to implement the mode sampling was to take into account
the distortions caused by different kinds of negative contributions by
enforcing the conservation laws\footnote{Remember that in
  event-by-event simulations inwards flow can be physical.}.

The probability distributions for the different quantities are rather
wide in the allcells sampling, so individual events in the more often
applied algorithm can have quantum numbers that are far away from
their actual values in that event. In a sense this is reasonable since
the equation of state used during the hydrodynamical stage, and the
particle distributions sampled at particlization assume a
grand-canonical ensemble where conservation laws are obeyed only in
average. But if we aim at a description of the collisions on an
event-by-event basis, we have a contradiction: In nature conservation
laws are obeyed in every single event. Thus it makes sense to require
that the conservation laws are strictly obeyed during all stages of
the model, and check how much the fluctuations created by the sampling
procedure affect the observables.

\subsection{Tests of Sampling Algorithm}
 \label{sec:tests}

\begin{figure}
\resizebox{0.5\textwidth}{!}{
  \includegraphics{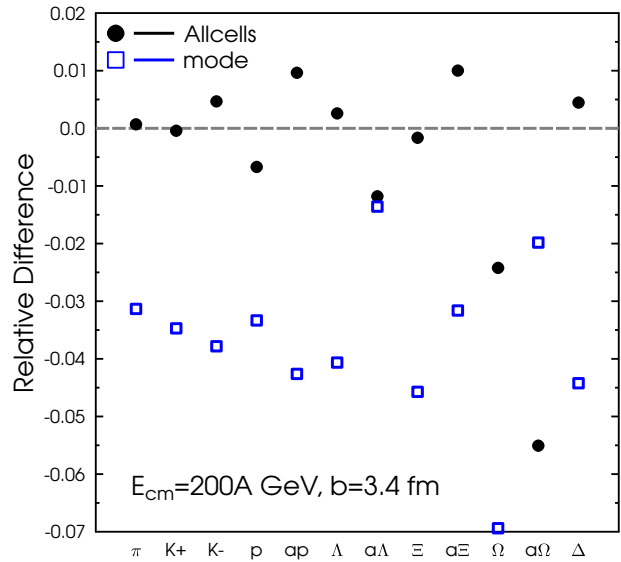}
}
\caption{Relative difference of sampled and integrated particle
  multiplicities in central Au+Au collisions at
  $E_{\rm cm}=200A$ GeV.}
\label{fig:nop_diff_rhic}       
\end{figure}

\begin{figure}
\resizebox{0.5\textwidth}{!}{
  \includegraphics{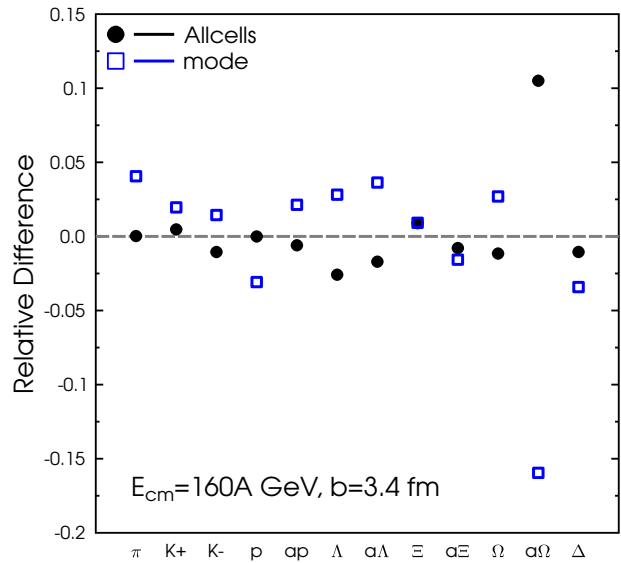}
}
\caption{Relative difference of sampled and integrated particle
  multiplicities in central Pb+Pb collisions at 
  $E_{\rm lab}=160A$ GeV.}
\label{fig:nop_diff_sps}       
\end{figure}

First of all, we need to check, whether the sampling algorithm
reproduces the numbers of particles for each species. In
Figs. \ref{fig:nop_diff_rhic} and \ref{fig:nop_diff_sps} the
multiplicities of selected particle species are compared to the
integrated results. Integration means here summing up $N_i$ as defined
in equation \ref{eqn_nop} for all hypersurface elements where
$\dif\sigma_\mu u^\mu >0$. All the results in this section are
comparisons of thermal yields for individual species only, the
resonance contribution has not been included.

\begin{figure}
\resizebox{0.5\textwidth}{!}{
  \includegraphics{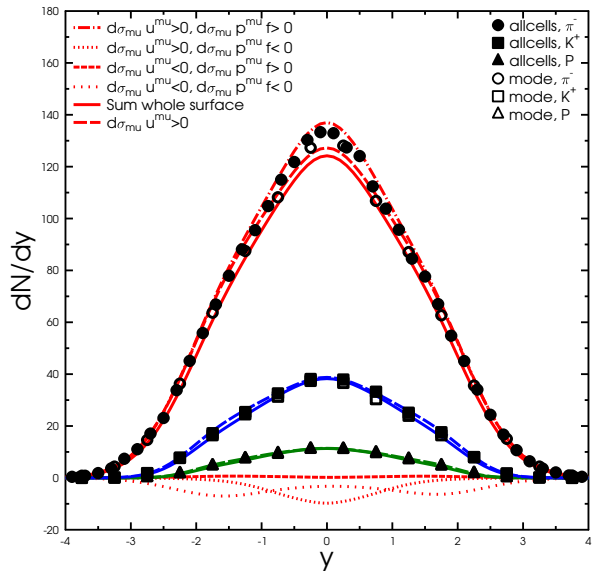}
}
\caption{Comparison of the integrated and sampled rapidity spectra for
  pions, kaons and protons in central Au+Au collisions at 
  $E_{\rm cm}=200A$ GeV.}
\label{fig:int_rap_rhic}       
\end{figure}

\begin{figure}
\resizebox{0.5\textwidth}{!}{
  \includegraphics{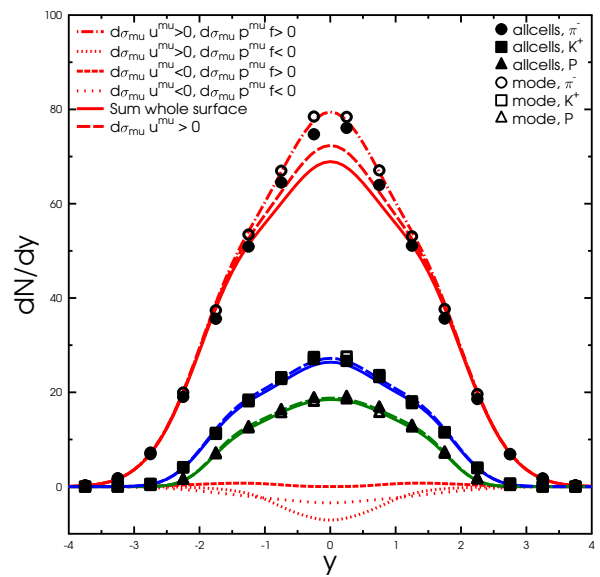}
}
\caption{Comparison of the integrated and sampled rapidity spectra for
  pions, kaons and protons in central Pb+Pb collisions at 
  $E_{\rm lab}=160A$ GeV.}
\label{fig:int_rap_sps}       
\end{figure}

We have plotted in Figs.~\ref{fig:nop_diff_rhic}
and~\ref{fig:nop_diff_sps} the relative deviations from the integrated
result, (\# sampled-\# integrated)/\# integrated, for the two sampling
algorithms. The deviations from zero are within the expected
statistical fluctuations for 500 events taking into account their
individual abundances for the allcells sampling.  Overall, in the
allcells sampling the particle multiplicities are nicely reproduced at
RHIC and SPS energies, whereas the mode sampling leads to slightly
less particles at RHIC. This result is expected, since we enforce a
lower value for the total energy of the system. The fluctuations of
the particle yields around that lower value are again purely
statistical. Since at SPS the inwards directed flow is smaller this
difference is also smaller and therefore the difference between
allcells and mode sampled yields is smaller. As one can see in the
rapidity spectra (Figs.~\ref{fig:int_rap_rhic}
and~\ref{fig:int_rap_sps}), at SPS energies mode sampling produces
more particles than allcells sampling.

The next step is to compare the distribution of the particles in
momentum space in terms of rapidity and transverse momentum. The
different lines in Fig. \ref{fig:int_rap_rhic} correspond to the
integrated result, which is split up into the different contributions
depending on the direction of flow and momentum with respect to the
particlization surface as described in the legend, see
Sec.~\ref{sec:neg}.

For clarity all the contributions are displayed only for pions while
for kaons and protons only the full result and the result from
hypersurface elements with flow directed outwards are shown. The full
line corresponds to the full result from the complete hypersurface
integration that one ideally wants to reproduce. The dashed line shows
the result for the contribution from the cells where the flow is
directed outward; these are the cells we take into account when
sampling the particle number densities. The momentum sampling on the
other hand disregards all the negative parts of the distribution
function. This treatment slightly skews the spectrum such that we end
up with a result that is in between the dashed and the dot-dashed line
for the sampled spectra. It may look surprising that the sampling does
not reproduce the yield from outward flowing elements at
midrapidity. After all, the sampling procedure has been normalised to
reproduce the yield from those elements. The reason for this deviation
is that the system in our calculations is not boost invariant. If it
were, the edges of the system in longitudinal direction would be
infinitely far away, and they would not contribute to midrapidity. But
in our non-boost invariant system they are close, and we see the same
to happen in longitudinal direction than what is depicted in
Fig.~\ref{fig:negpt} for transverse direction: The low momentum, \ie
small rapidity, region is depleted because of the negative
contributions. The distributions we sample do not contain this
depletion, and thus we end up having too many particles at
midrapidity. On the other hand, since the sampling procedure
reproduces the total yield of particles as shown in
Figs.~\ref{fig:nop_diff_rhic} and~\ref{fig:nop_diff_sps}, we must have
fewer particles somewhere. A closer look at the spectra reveals that
the sampled yield is below the $\dif\sigma_\mu u^\mu > 0$ curve at
rapidities $1<|y|<2.5$, and this deficit is sufficient to cover the
excess at midrapidity.

\begin{figure}
\resizebox{0.5\textwidth}{!}{
  \includegraphics{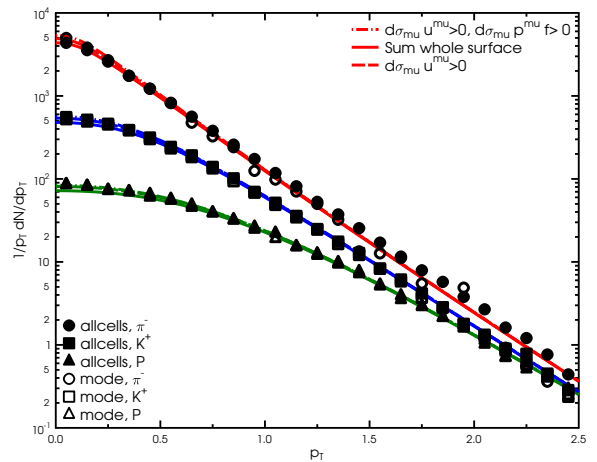}
}
\caption{Comparison of the integrated and sampled transverse momentum
  spectra for pions, kaons and protons in central Au+Au collisions at
  $E_{\rm cm}=200A$ GeV.}
\label{fig:int_pt_rhic}       
\end{figure}

The transverse momentum spectrum is shown in
Fig. \ref{fig:int_pt_rhic}. Due to the logarithmic scale in this case,
the differences shown in Fig~\ref{fig:negpt} disappear, and only a
small difference between the elements with flow directed outwards and
the full result is visible at low transverse momentum. The sampled
results are in very good agreement with the integrated results. All
possible deviations (apart from statistical noise) are smaller than
the size of the symbols.

\begin{figure}
\resizebox{0.5\textwidth}{!}{
  \includegraphics{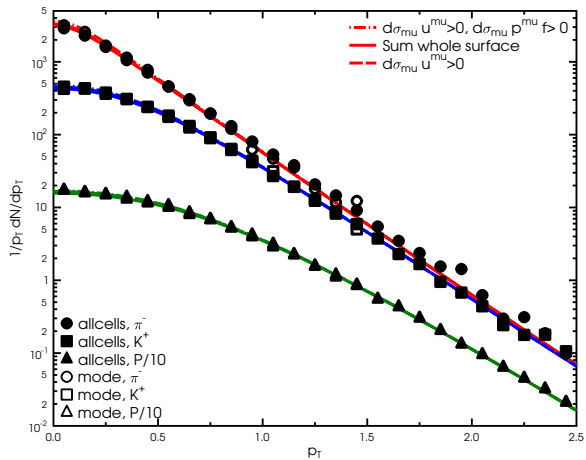}
}
\caption{Comparison of the integrated and sampled transverse momentum
  spectra for pions, kaons and protons in central Pb+Pb collisions at
  $E_{\rm lab}=160A$ GeV.}
\label{fig:int_pt_sps}       
\end{figure}

The comparison of rapidity spectra and transverse momentum spectra at
SPS, as shown in Figs. \ref{fig:int_rap_sps} and \ref{fig:int_pt_sps},
leads to very similar results. The only difference is that in the mode
sampling there are more pions produced than in the allcells sampling
as mentioned above. The reason for this has to be investigated further
in the future.

Overall, the sampling algorithms reproduce the particle yields and
spectra very well within the expected deviations from the assumptions
that are made in the different algorithms. The negative contributions
are also smaller than 5 \% in the regions where most of the particles
are emitted and affect heavier particles less than lighter ones.

\subsection{Effects of Rescattering}

\begin{figure}
\resizebox{0.5\textwidth}{!}{
  \includegraphics{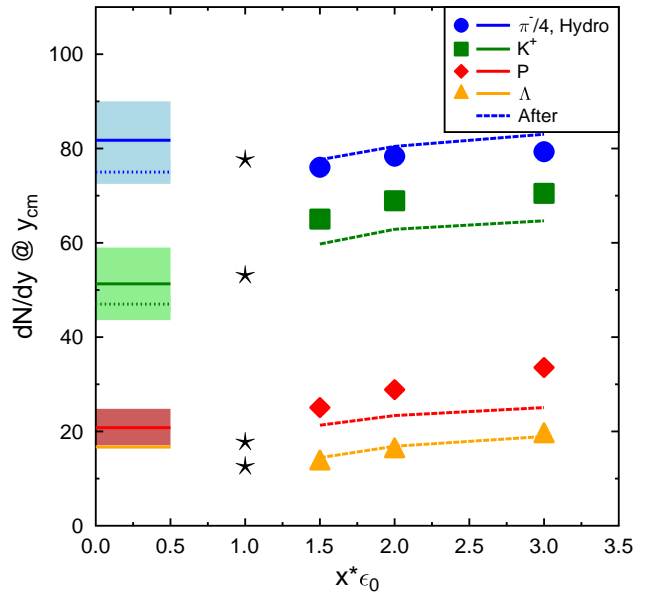}
}
\caption{Yields at midrapidity ($|y|<0.5$) as a function of switching
  criterion in multiples of $\epsilon_0$ for four different particle
  species ($\pi^-$, $K^+$, P, $\Lambda$) in central ($b<3.4$ fm) Au+Au
  collisions at $\sqrt{s_{\rm NN}}=200$ GeV. The symbols indicate the
  result with resonance decays ('Hydro') and the lines show the result
  after rescattering ('After'). The black stars show the result from
  the previously used 'gradual' transition. The bands on the left
  indicate data by the STAR and BRAHMS collaborations
  \cite{Ouerdane:2002gm,Lee:2004bx,Adams:2003xp,Adams:2006ke}. }
\label{fig:mul_midy_rhic}       
\end{figure}

In this Section we present selected results for multiplicities,
spectra and elliptic flow to demonstrate the effect of the hadronic
rescattering and the switching criterion. All the results are
calculated using the hybrid approach described in
Sec.~\ref{sec:hybrid} with mode sampling enforcing exact conservation
laws. To obtain the results directly at the switching hypersurface we
run UrQMD for $0.1$ fm and use it only to calculate the resonance
decays ('Hydro'). The second option is to run UrQMD for $200$ fm and
include all the rescattering dynamics in the hadron transport approach
('After').

\begin{figure}
\resizebox{0.5\textwidth}{!}{
  \includegraphics{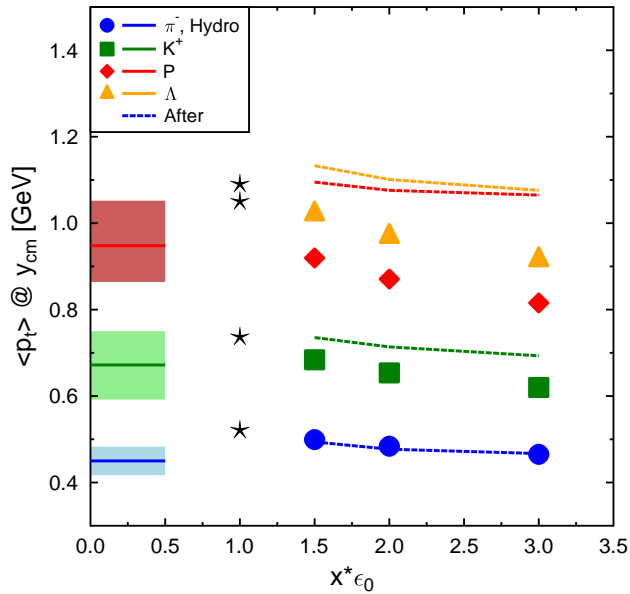}
}
\caption{Mean transverse momentum at midrapidity ($|y|<0.5$) as a
  function of switching criterion in multiples of $\epsilon_0$ for
  four different particle species ($\pi^-$, $K^+$, P, $\Lambda$) in
  central ($b<3.4$ fm) Au+Au collisions at $\sqrt{s_{\rm NN}}=200$
  GeV. The symbols indicate the result with resonance decays ('Hydro')
  and the lines show the result after rescattering ('After'). The
  black stars show the result from the previously used 'gradual'
  transition. The bands on the left indicate data by the PHENIX
  collaboration~\cite{Adler:2003cb}. }
\label{fig:mpt_midy_rhic}       
\end{figure}

In Fig.~\ref{fig:mul_midy_rhic} the yields of pions, kaons, protons
and Lambdas are shown in central heavy ion collisions at the highest
RHIC energy. The final results using allcells sampling are within 3\%
to the ones shown here (see Fig. \ref{fig:nop_diff_rhic}). Therefore,
we decided to restrict the number of shown results to mode sampling
only. The switching criteria that we have chosen correspond to the
following temperatures $3 \epsilon_0 \approx 163$ MeV, 
$2 \epsilon_0 \approx 154$ MeV and $1.5 \epsilon_0 \approx 149$ MeV
and are in the ball-park of switching criteria that are commonly used
in other hybrid
approaches~\cite{Bass:2000ib,Teaney:2001av,Hirano:2005xf,Nonaka:2006yn,Petersen:2008dd,Werner:2010aa,Song:2010mg,Karpenko:2012yf,Hirano:2012kj}.
The energy density $\epsilon_0$ refers to the nuclear ground state
energy density 146 MeV/fm$^3$. The switching criterion needs to be
adjusted together with equation of state and parameters defining the
initial conditions to achieve a good agreement with experimental
data. In this analysis we concentrate on presenting general features
of the results and have not searched for the best agreement with the
measurements yet.

\begin{figure}
\resizebox{0.5\textwidth}{!}{
  \includegraphics{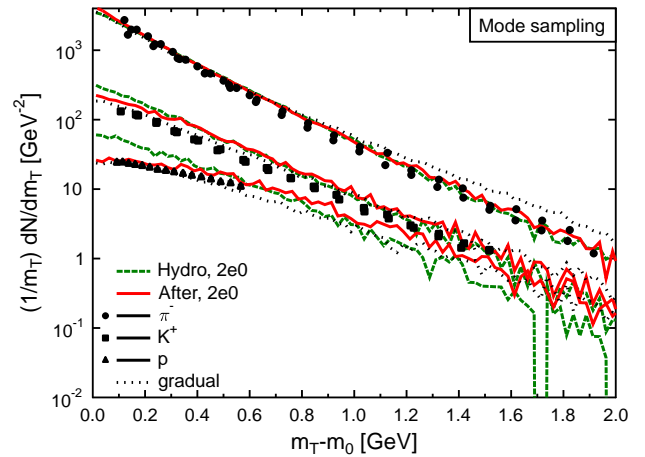}
}
\caption{Transverse mass spectra at midrapidity ($|y|<0.5$) for pions,
  kaons and protons in central ($b<3.4$ fm) Au+Au collisions at
  $\sqrt{s_{\rm NN}}=200$ GeV. The dashed lines indicate the result
  with resonance decays ('Hydro') and the full lines show the result
  after rescattering ('After'). The black dotted line represents the
  result from the previously used 'gradual' transition. The symbols
  indicate experimental data by the STAR, PHENIX and BRAHMS
  collaborations~\cite{Adams:2003xp,Arsene:2005mr,Adler:2003cb,nucl-ex/0601042}.
  The proton data by STAR has been multiplied by 0.6 to correct for
  the feed-down from Lambdas.}
\label{fig:dndmt_midy_rhic}       
\end{figure}

The yields are slightly higher for higher switching energy densities,
because in equilibrium at this temperature range, higher temperatures
lead to higher yields. The deviations from equilibrium caused by the
sampling procedure are not large enough to break this rule. The
rescattering leads to a reduction of the kaon and proton yield,
whereas pions and Lambdas are not so much affected. The kaon and
proton yields are also higher in the iso-energy density transition
scenario than in the previously employed gradual transition, where
full transverse slices are sampled on an isochronous surface, when the
whole slice has diluted below 5$\epsilon_0$. Apart from the fact that
the kaon yield is higher than the experimental data our results are in
reasonable agreement with the data.

\begin{figure}
\resizebox{0.5\textwidth}{!}{
  \includegraphics{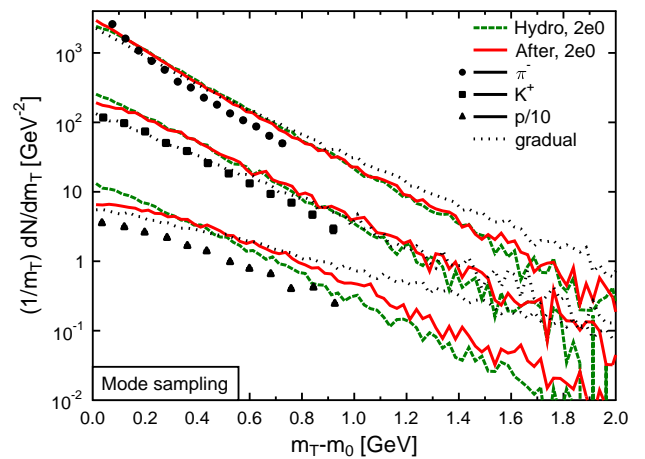}
}
\caption{Transverse mass spectra at midrapidity ($|y|<0.5$) for pions,
  kaons and protons in central ($b<3.4$ fm) Pb+Pb collisions at
  $E_{\rm lab}=160A$ GeV. The dashed lines indicate the result with
  resonance decays ('Hydro') and the full lines show the result after
  rescattering ('After'). The black dotted line represents the result
  from the previously used 'gradual' transition. The symbols indicate
  experimental data by the NA49
  collaboration~\cite{Afanasiev:2002mx,Alt:2006dk}.}
\label{fig:dndmt_midy_sps}       
\end{figure}

The mean transverse momentum as a measure of the radial flow
development is shown in Fig. \ref{fig:mpt_midy_rhic}. In this case the
dependence on the switching criterion is opposite to the one of the
yields. There are two reasons for this behaviour: The first is that
for lower switching energy densities the system stays longer in the
hydrodynamic evolution and the particles gain more transverse
flow. The second argument is that the total energy needs to be
conserved and if there are more particles produced, there is less
energy remaining to give them kinetic energy in terms of transverse
flow at the switching transition. Depending on their hadronic
cross-sections and their mass the particles get pushed a lot
(e.g. protons) or almost not at all (pions) during the hadronic
rescattering stage.

\begin{figure}
\resizebox{0.5\textwidth}{!}{
  \includegraphics{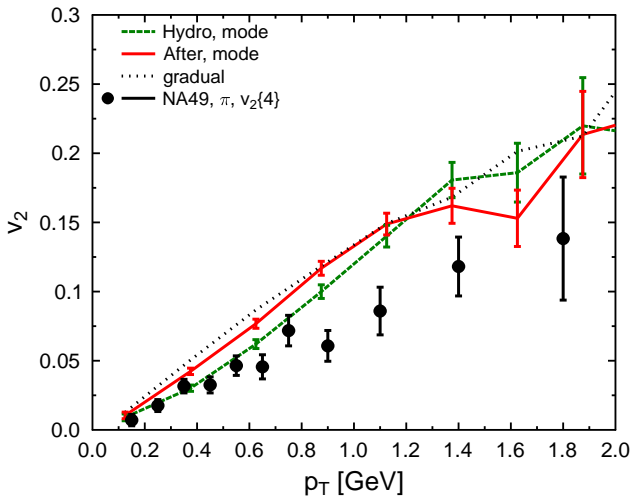}
}
\caption{Elliptic flow of pions as a function of transverse momentum
  at midrapidity ($|y|<0.5$) in mid-central ($b=7$ fm) Pb+Pb
  collisions at $E_{\rm lab}=160A$ GeV. The dashed lines indicate the
  result with resonance decays ('Hydro') and the full lines show the
  result after rescattering ('After'). The black dotted line
  represents the result from the previously used 'gradual'
  transition. The symbols indicate experimental data by the NA49
  collaboration \cite{Alt:2003ab}.}
\label{fig:v2_ptpi_sps}       
\end{figure}

The transverse mass spectra at RHIC and SPS (shown in
Figs. \ref{fig:dndmt_midy_rhic} and \ref{fig:dndmt_midy_sps}
respectively) confirm this behaviour. The pion spectra are almost
identical before and after the hadronic cascade, but the kaon and
proton spectra begin to resemble the experimental data only after the
rescattering. The comparison to the previously imposed gradual
transition shows that a true iso-energy density switching criterion
improves the slope of the spectra at high transverse masses
drastically. This can be easily understood, since in the gradual
transition scenario the full slice needs to reach the energy density
criterion which delays the switching to hadron cascade for the edges
in the transverse direction. The edges gain very large transverse flow
velocity, which makes the distributions flatter, and even if the edges
get very cold, the yield of heavy particles does not drop so much that
the effect of large flow velocity would be negated. What comes to the
$p_T$-distributions, the hybrid model based on ideal hydrodynamics
works better at high RHIC energies than at high SPS energies. This
indicates that the non-equilibrium dynamics gains importance at lower
beam energies.

\begin{figure}
\resizebox{0.5\textwidth}{!}{
  \includegraphics{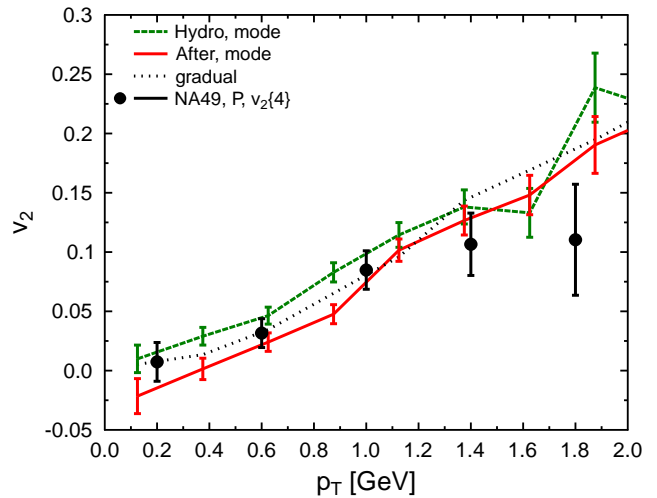}
}
\caption{Elliptic flow of protons as a function of transverse momentum
  at midrapidity ($|y|<0.5$) in mid-central ($b=7$ fm) Pb+Pb
  collisions at $E_{\rm lab}=160A$ GeV. The dashed lines indicate the
  result with resonance decays ('Hydro') and the full lines show the
  result after rescattering ('After'). The black dotted line
  represents the result from the previously used 'gradual'
  transition. The symbols indicate experimental data by the NA49
  collaboration \cite{Alt:2003ab}.}
\label{fig:v2_ptp_sps}       
\end{figure}

In Figs.~\ref{fig:v2_ptpi_sps} and \ref{fig:v2_ptp_sps} elliptic flow
as a function of transverse momentum is shown for pions and protons at
high SPS energy. Quite surprisingly the $p_T$-differential elliptic
flow is quite insensitive to the switching criterion. It is known that
in ideal fluid hydrodynamics pion $v_2(p_T)$ is quite insensitive to
the freeze-out temperature~\cite{Kolb:2000}. We presume that the same
holds also when one uses the gradual criterion instead of constant
temperature/density for particlization, and thus the elliptic flow of
pions at particlization is very similar in both cases, and evolves
similarly during the cascade. Thus the final $v_2(p_T)$ is similar as
well. The proton $v_2(p_T)$ shows more sensitivity than pion
$v_2(p_T)$, but the difference is at low $p_T$, not at high $p_T$
which is influenced by the different evolution of the edges as
explained when discussing the $p_T$ spectra. On the other hand, the
lower proton $v_2$ at low $p_T$ is in line with the ideal fluid
expectations where lower freeze-out temperature leads to lower proton
$v_2(p_T)$ at low $p_T$. Thus the low $p_T$ range of protons is mostly
influenced by the center of the system which evolves hydrodynamically
much longer when one uses isodensity switching than when one uses
gradual switching. When the hadronic cascade rescattering is taken
into account, the proton flow at low transverse momenta seems to even
turn negative as recently observed by the CERES
collaboration~\cite{Adamova:2012md}.

\section{Conclusions}

In this paper we have discussed a crucial part of hybrid models in
detail: How to switch from hydrodynamical description to cascade. We
have described an algorithm to find an isovalue surface where the
transition takes place, the unavoidable negative contributions of
Cooper-Frye procedure on such a surface, and how to make the sampling
algorithm, which creates the initial state for the cascade, to work on
an arbitrary surface and deal with the negative contributions. We have
seen that in realistic calculations the negative contributions are not
a large problem, but they create an uncertainty of their own which
should be kept in mind when one draws conclusions from the results of
the present hybrid or hydrodynamical models. In the long run we hope
to develop a model where these negative contributions are properly
treated, and not just ignored.

We think that the machinery we have described is particularly
suitable for a proper event-by-event analysis of heavy-ion
collisions. In event-by-event calculations the initial state of
hydrodynamics varies wildly from event to event, and thus one may
expect the particlization surface to have a very complicated
structure. This poses no problem for the algorithm described here
since it can create consistent surfaces without holes nor double
counting for any configuration. One of the reasons for studying
heavy-ion collisions event-by-event are fluctuations. In those studies
it is important to distinguish at which stage of the evolution
fluctuations are created and how. The sampling algorithm we described
provides an important check by allowing strict conservation of energy
and charges thus guaranteeing that their fluctuations are not caused
by the sampling procedure.

All this machinery is not limited to switching from ideal fluid to
cascade, but it can be applied at the interface of viscous
hydrodynamics and cascade as well. All one needs to do is to modify
the particle distributions at particlization accordingly.

However, we presented here preliminary results only without attempting
to describe the data well. The next step is to identify the main
parameters of the hybrid calculation and perform a multi-parameter
analysis compared to bulk observables at RHIC using a sophisticated
statistical emulator to limit CPU time. Once a good switching
criterion has been identified, the beam energy dependence can be
explored. In this context, the framework presented here can easily be
even more generalised to other switching criteria based on net baryon
density, temperature or combinations of thermodynamic quantities.

\section*{Acknowledgements:}
We are greatly indebted to Bernd Schlei for many informative
discussions about surface finding and related algorithms. We thank
Tetsufumi Hirano, Huichao Song and Hannu Holopainen for describing
their sampling routines in detail. The work of P.H.\ is supported by
BMBF under contract no.\ 06FY9092 and the work of H.P.\ by U.S.\
Department of Energy grant DE-FG02-05ER41367. We are grateful to the
Open Science Grid for providing computing resources. H.P.\ thanks
J.~W.~Goethe Universit\"at and HICforFAIR for hospitality during
several stages of this work.

\end{document}